\documentclass[universe,article,accept,pdftex,moreauthors]{Definitions/mdpi} 
\firstpage{1} 
\pubvolume{11}
\issuenum{11}
\articlenumber{370}
\pubyear{2025}
\copyrightyear{2025}
\usepackage{Definitions/aas_macros}

\externaleditor{Marco Berton} 
\datereceived{15 August 2025} 
\daterevised{3 November 2025} 
\dateaccepted{5 November 2025} 
\datepublished{7 November 2025} 
\hreflink{https://doi.org/\linebreak 10.3390/universe11110370} 

\Title{Probing Jet Compositions with Extreme Mass Ratio  Binary\linebreak   Black Holes}

\TitleCitation{Probing Jet Compositions with Extreme Mass Ratio  Binary Black Holes}

\Author{{Hung-Yi Pu} 
 $^{1, 2, 3}$\orcidA{}}

\AuthorNames{Hung-Yi Pu}

\isAPAStyle{%
       \AuthorCitation{Hung-Yi Pu}
         }{%
        \isChicagoStyle{%
        \AuthorCitation{Lastname, Firstname, Firstname Lastname, and Firstname Lastname.}
        }{
        \AuthorCitation{Pu, H.-Y.}
        }
}

\address{%
$^{1}$ \quad Department of Physics, National Taiwan Normal University, No. 88,  Section 4, Tingzhou Road,\linebreak   Taipei {116,} 
 Taiwan; hypu@gapps.ntnu.edu.tw\\
$^{2}$ \quad Center of Astronomy and Gravitation, National Taiwan Normal University, No. 88,  Section 4, Tingzhou Road, Taipei {116,} Taiwan\\
$^{3}$ \quad Institute of Astronomy and Astrophysics, Academia Sinica, 11F of Astronomy-Mathematics Building, AS/NTU No. 1, {Section} 
 4, Roosevelt {Road}, Taipei {10617,} Taiwan}

\abstract{Determining whether black hole jets are dominated by leptonic or baryonic matter remains an open question in high-energy astrophysics. We propose that extreme mass ratio   binary (EMRB) black holes, where an intermediate mass secondary black hole (a ``miniquasar'') periodically interacts with the accretion flow of a supermassive black hole (SMBH), offer a natural laboratory to probe jet composition. 
{{In } 
 an  EMRB, the miniquasar jet  is   launched episodically after each disk-crossing event, triggered by the onset of super-Eddington accretion. 
The resulting emissions exhibit temporal evolution as the jet interacts with the SMBH accretion disk. Depending on whether the jet is leptonic or hadronic in composition, the radiative signatures differ substantially. Notably, a baryonic jet produces a more pronounced gamma-ray output than a purely leptonic jet}.
By modeling the evolution of  { the multifrequency characteristic} features, { it is suggested that the gamma-ray-to-UV emissions may serve as a diagnostic tool} capable of { distinguishing}  between leptonic and baryonic scenarios. The resulting electromagnetic signals, when combined with multi-messenger observations, offer a powerful means to constrain the physical nature of relativistic jets from black holes.}


\begin{document}

\section{Introduction}
Relativistic jets from black hole systems are observed across a wide range of mass scales, from stellar-mass black holes, known as microquasars {e.g.,} 
~\citep[]{Fender2004,cor2011,kylasfis2012}, to supermassive black holes that power active galactic nuclei (AGNs) e.g.,~\citep[]{trump2011,fabian2012,Hov2019,combes2021}. AGN jets can extend over vast distances, sometimes spanning thousands to millions of light years, and have significant energy and momentum, capable of shaping their surrounding environments~ \citep[][]{asada2012,kra2025,fabian2012}. Together with complicated jet acceleration e.g.,~\citep[]{asada2014} and particle acceleration within the jet e.g.,~\citep[]{marin2024}, the jet composition, whether dominated by leptons or hadrons, plays a crucial role in determining how energy is transported, dissipated and transferred. Due to their differing masses and interaction cross sections with ambient media such as intracluster medium, leptons and hadrons affect the jet's energy dissipation and acceleration mechanisms in fundamentally different ways~\citep[]{fan2018, owen2022_lep,owen2022}.
In addition, the composition of the cosmic ray can also be related to the particles available within the jet and experience further acceleration ~\citep[]{Matteo2020,lin2023,kan2023}. 

The spectral properties of AGNs, particularly blazars e.g.,~\citep[]{Hov2019,Matteo2020,Rodrigues2024}, provide information on the composition of the jet. A typical blazar spectrum displays two distinct peaks. It is widely accepted that leptons produce the lower-frequency peak through synchrotron radiation \citep{Hov2019}. However, the source of the higher-frequency peak at  gamma-ray range remains unclear. This ambiguity is due to the possibility that gamma-ray emissions could either result from inverse Compton (IC) scattering by leptons or through specific hadronic processes, such as the decay of pions leading to gamma-ray production.

Compared to the jets from steady-state, isolated sources, jets originating from black hole systems might offer valuable insights into jet composition, for example, time-varying systems with episodic, transient jets from binary systems, such as OJ 287 \cite[][]{lico2022,val2024}, GRO 1655-40 \cite{zheng1997}, XTE J1550-564 \cite{kaa2003}, or systems where jets interact with the surrounding matter~\citep[]{owen2022_lep,owen2022}.
Motivated by this paradigm, we investigate the emission characteristics of periodic jets in extreme mass ratio  binary black holes (EMRBs). We focus on  EMRBs  with a mass ratio $q=M_{\rm secondary}/M_{\rm primary}<10^{-4}~{\rm to}~10^{-6}$ consisting of an SMBH with $M_{\rm SMBH}=10^{6}~{\rm to}~10^{9} {\rm M}_{\odot}$ as the primary and a secondary black hole with\linebreak   $M_{\rm mq}=10^{2}~{\rm to}~10^{5} {\rm M}_{\odot}$ (hereafter, we refer to such intermediate mass black holes as ``miniquasars''). 

\begin{figure}[H]
\includegraphics[width=0.8\textwidth]{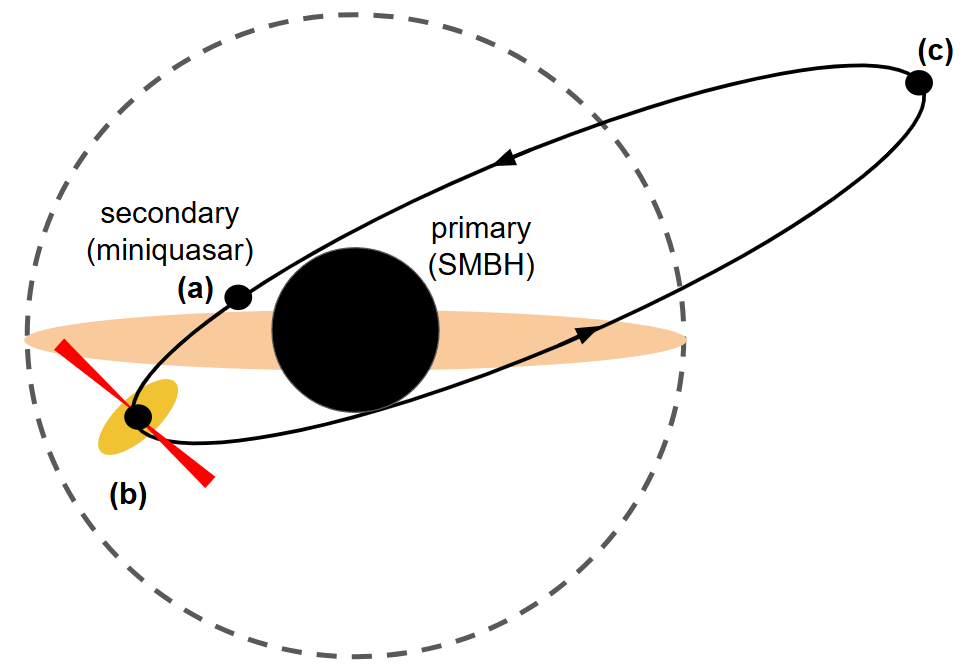}
\caption{
{A} 
 schematic illustration of an  EMRB considered here (not to scale). The system comprises a primary SMBH surrounded by a thin accretion disk and a secondary intermediate mass ``miniquasar'' in an eccentric elliptical orbit. Periodic interactions between the miniquasar and the SMBH accretion flow produce distinct orbital phases: (\textbf{a}) the miniquasar approaches pericenter, prior to interacting with the SMBH's disk; (\textbf{b}) following the collision, the miniquasar begins accreting captured material, launching a relativistic jet that interacts with the SMBH's accretion environment; (\textbf{c}) as the miniquasar moves toward the apocenter, the captured material is exhausted and accretion stops. This sequence of phases (\textbf{a})--(\textbf{c}) repeats over each orbit. For reference, the dashed circle marks the approximate outer edge of the SMBH accretion disk ($\sim$$10^{3}\,R_{\rm g}$).
\label{fig:overview}}
\end{figure}

 These EMRB systems provide a promising avenue for probing the composition of relativistic jets through their associated radiative signatures, since transient jet-launching and jet–ambient interactions naturally occur periodically. 
Figure \ref{fig:overview} displays a schematic plot for the  EMRB of interest, in which the semi-major axis extends beyond the boundary of the SMBH disk ($\sim$$10^3 R_{\rm g}$, where $R_{\rm g}\equiv GM_{\rm SMBH}/c^2$). The configuration enables periodic interactions between the miniquasar and the thin accretion disk around the SMBH, and the lauching and quenching of the miniquasar jet. 
As the miniquasar traverses the SMBH accretion flow (between phases (a) and (b) in Figure \ref{fig:overview}), it can capture material from the SMBH accretion disk. This captured mass is subsequently accreted by the miniquasar, thus initiating the miniquasar jet (phase (b) in Figure \ref{fig:overview}). Before the miniquasar reaches the apocenter, the mass is already depleted, leading to the cessation of the jet (phase (c) in Figure \ref{fig:overview}).  
The transient accretion considered here is different from the continuous accretion when the secondary is embedded in the SMBH accretion disk due to capture by dynamical interactions of nuclear star clusters within the AGN disk e.g.,~\citep[]{tagawa2023a,tagawa2023}, or because the secondary orbits the AGN with an  untilted circular orbit during the inspiral phase \citep{kocsis2011}.  In addition, the emission features due to the encounter between the accretion disk of an SMBH with another SMBH are studied in ~\citep[]{Lehto1996,pihajoki2016}.

 In this work, we specifically focus on emissions resulting from the interaction between the miniquasar jet and the SMBH accretion disk (as illustrated by  phase (c) in Figure \ref{fig:overview}) and its evolution. By examining the orbital-phase-dependent multi-wavelength emissions during the jet--disk collision, we demonstrate how the light curves of the UV and gamma-ray power would vary with different jet compositions.  The inherent orbital modulation in these binaries offers a unique advantage over jets from isolated AGNs or standalone miniquasars,  breaking the spectral degeneracies that typically hinder efforts to constrain jet composition in steady-state observations.

 For  EMRBs  with SMBH $M_{\rm SMBH}$ mass and semi-major axis $a$, the orbital {period is} 
 
\begin{equation}
    T=2\pi\sqrt{\dfrac{a^{3}}{GM_{\rm SMBH}}}\;.
\end{equation}

\begin{figure}[H]
\includegraphics[width=0.8\textwidth]{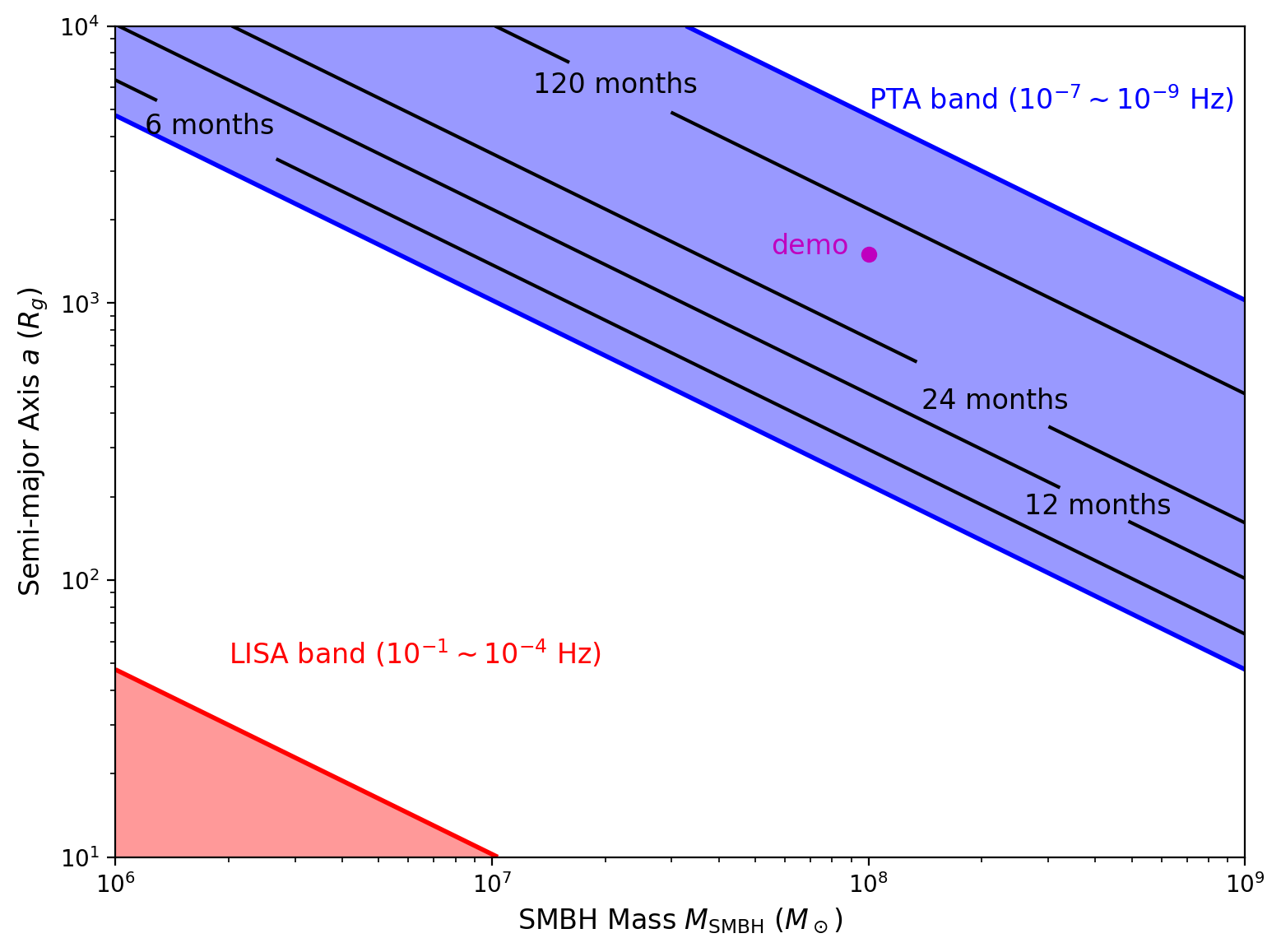}
\caption{{Orbital} 
 period as function of semi-major axis $a$ and the mass of the SMBH, in months. The sensitive frequency ranges for PTA and LISA are indicated by the blue and red regions. The selected parameters for our demonstrative case is indicated by the magenta dot.
 \label{fig:period}}
\end{figure}

{The} 
 orbital period for  EMRBs  with different $M_{\rm SMBH}$ and $a$ is presented in \mbox{Figure \ref{fig:period}.}
 The orbital frequencies of the  EMRBs  of interest here, with its apocenter outside the edge of the SMBH disk and periods typically on the order of years, lie within the sensitivity band of the pulsar timing arrays (PTAs) \citep{mclaughlin2013,kramer2013,manchester2013,verbiest2016}, as indicated by the blue-shaded region in \mbox{Figure \ref{fig:period}.}  The EMRBs considered here are the progenitors of extreme mass ratio inspiral (EMRI) systems. With AGN-assist dynamics and the decay of orbital energy by gravitational waves, the systems gradually spiral and transition into EMRIs \citep[][]{pna2021}, eventually reaching higher frequencies detectable by the LISA \citep{klein2016,amaro2017}, as indicated by the \mbox{red-shaded region.}

\vspace{-12pt}

The paper is organized as follows. 
In Section \ref{S2}, we provide an overview of the leptonic and hadronic composition of jets, and related radiation mechanisms. In Section \ref{S3}, we present our method and results for modeling the phased resolved jet emission features, including the computation of orbits and estimation of leptonic and hadronic emissions from the jet. The summary and final remarks are provided in Section \ref{S4}.

\section{Jet Composition and Radiation Mechanism}\label{S2}
\subsection{Jet Composition} \label{sec: jet_composition}
Based on the observed synchrotron emissions below gamma-ray \citep[][]{Hov2019,Matteo2020}, electrons are identified as an essential component of the jet composition. 
The number density ratio between different species, electrons ${\rm e^{-}}$, positron ${\rm e^{+}}$, and protons ${\rm p}$  are denoted by
\begin{equation}\label{eq:xi}
\begin{array}{cc}
    \xi_{\rm e^{+}}&\equiv \dfrac {n_{ \rm e^{+}}}{n_{ \rm e^{-}}}\;,\\
    \xi_{\rm p}&\equiv \dfrac {n_{ \rm p}}{n_{\rm e^{-}}}\;,
\end{array}
\end{equation}
{and} charge neutrality require
\begin{equation}\label{eq:constraint:xi}
   \xi_{\rm e^{+}}+\xi_{\rm p}=1\;. 
\end{equation}

The total  intrinsic jet power $P_{\rm jet}$ can therefore be decomposed into powers carried by electrons $P_{\rm e^{-}}$, positron $P_{\rm e^{+}}$, magnetic field $P_{\rm B}$, and protons $P_{\rm p}$:
\begin{equation}\label{eq:Ltotal}
P_{\rm jet}=P_{\rm e^{-}}+P_{\rm e^{+}}+P_{\rm p}+P_{\rm B}\;.
\end{equation}

{After} normalizing the decomposed power with {$P_{\rm e^{-}}$} 
 by
\begin{equation}\label{eq:L_ratio}
\begin{array}{cl}
    P_{\rm e^{+}}&=\xi_{\rm e^{+}}P_{\rm e^{-}}\;,\\
      P_{\rm p}&=\dfrac{{\rm m}_{p}\langle\gamma_{\rm p}\rangle}{{\rm m}_{\rm e}\langle\gamma_{\rm e}\rangle}\xi_{\rm p}P_{\rm e^{-}}\;,  \\
    P_{\rm B}&=P_{\rm e^{-}}+P_{\rm e^{+}}\;,
\end{array}
\end{equation}

The total jet power, Equation (\ref{eq:Ltotal}), can therefore be rewritten as
\begin{equation}
P_{\rm jet}=P_{\rm e^{-}}\left(2+2\xi_{\rm e^{+}}+\dfrac{{\rm m}_{\rm p}\langle\gamma_{\rm p}\rangle}{{\rm m}_{\rm e}\langle\gamma_{\rm e}\rangle} \xi_{\rm p}\right)\;.
\end{equation}

{In} the above formula, $m_{\rm p}/m_{\rm e}(\approx$$1836$) is the mass ratio between a proton and an electron, $\langle\gamma_{\rm e}\rangle$ and $\langle\gamma_{\rm p}\rangle$ are the average particle Lorentz factors for electrons and positrons and protons in the jet comoving frame, respectively, and the equipartition between magnetic energy and lepton energy is adopted. 
 Assuming the emissions of the jet iarecharacterized by a power-law spectrum with power-law index $\bar{\alpha}$, with the view angle $\theta$ and jet bulk Lorentz factor $\Gamma=(1-\beta^{2})^{-1/2}$, where $\beta$ is the ratio of the speed of the jet and the speed of light, the observed jet power $P_{\rm jet}^{\rm obs}$ is related to the intrinsic jet power by 
\begin{equation}\label{eq:Pjet_obs}
    P_{\rm jet}^{\rm obs}=\delta^{3+\bar{\alpha}} P_{\rm jet}\;,
\end{equation}
with the Doppler factor 
\begin{equation}\label{eq:doppler_factor}
    \delta=\frac{1}{\Gamma(1-\beta \cos\theta)}\;.
\end{equation}

Furthermore, the power ratio of different species can be defined by 
\begin{equation}\label{eq:eta_e}
    \begin{array}{cl}
    \eta_{\rm e^{-}}&\equiv\frac{\rm P_{\rm e^{-}}}{\rm P_{\rm jet}},
\\
    \eta_{\rm e^{+}}&\equiv\frac{\rm P_{\rm e^{+}}}{\rm P_{\rm jet}}=\xi_{\rm e^{+}}\eta_{\rm e^{-}},
\\
    \eta_{\rm p}&\equiv\frac{\rm P_{\rm p}}{\rm P_{\rm jet}}=\dfrac{{\rm m}_{\rm p}\langle\gamma_{\rm p}\rangle}{{\rm m}_{\rm e}\langle\gamma_{\rm e}\rangle} \xi_{\rm p} \eta_{\rm e^{-}}.
\end{array}
\end{equation}

\begin{figure}[H]
\includegraphics[width=0.9\textwidth]{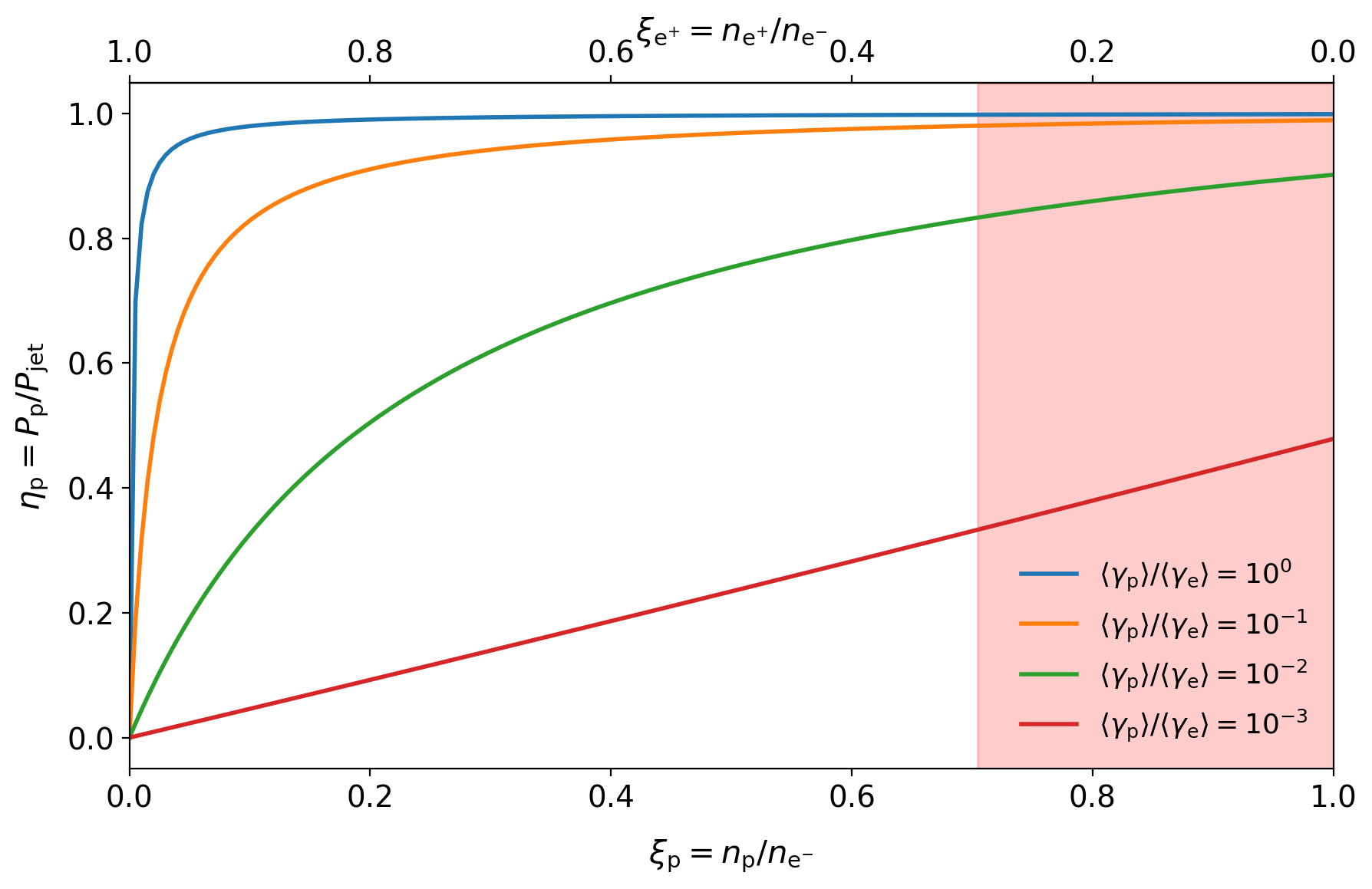}
\caption{{Fraction} 
 of the jet power carried by hadronic  components, $\eta_{{\rm p}}$, for varying ratios between the particle Lorentz factor of proton and leptons, $\langle\gamma_{\rm p}\rangle/\langle\gamma_{\rm e}\rangle$; see Equation (\ref{eq:eta_e}). The shaded region corresponds to cases in which the baryonic component accounts for more than 99.9$\%$ of the \mbox{total mass}.\label{fig:eta_compare}}
\end{figure}

{Therefore, the} energy content of the jet depends on the kinetic energy ratio between baryons and leptons $\langle\gamma_{\rm p}\rangle/\langle\gamma_{\rm e}\rangle$. Figure \ref{fig:eta_compare} displays the relative contribution of the hardronic components of the jet power,  $\eta_{\rm p}$, with different ratios of $\langle\gamma_{p}\rangle/\langle\gamma_{e}\rangle$. The higher $\langle\gamma_{\rm p}\rangle/\langle\gamma_{\rm e}\rangle$ or $\xi_{p}$, the more jet power is carried by the baryons.

Based on the mass content of the species within the jet, three potential configurations can be classified:
\begin{itemize}
    \item {\bf {Pair-Dominated/Leptonic Jet}
}: The mass of a pair-dominated jet is dominated by electrons and positrons, with a negligible contribution from protons $\xi_{\rm p}\simeq0$. The efficiency of leptonic loading into jets can be related to pair loading $\gamma+\gamma\to e^{-}+e^{+}$ with MeV photons \citep[][]{Monika2011,wong2021,kimura2022}.
    \item {\bf{Baryonic Jet}}: A baryonic jet's mass is dominated by its baryonic (hadronic) component: protons and/or heavier nuclei. For example, 
    the shaded region in \mbox{Figure \ref{fig:eta_compare}} indicates a case when the barynoic component contributes $>99.9\%$ of the total mass, and a representative baryonic jet. A commonly adopted  pure electron--proton jet with $\xi_{\rm p}=1$  belongs to this model. 
    In a baryonic jet,  although the mass is predominantly composed of baryonic matter, the energy content may not be similarly dominated. This occurs when leptons have significantly higher energy $\langle\gamma_{\rm e}\rangle\gg \langle\gamma_{\rm p}\rangle$, as illustrated in Figure \ref{fig:eta_compare}.  
    The efficient acceleration of the jet's hadronic components is crucial for generating ultra-high-energy cosmic rays (UHECR) \citep[][]{globus2023} and UHE \mbox{neutrinos \citep[][]{meszaros2017}.}
    Neutrino-producing blazars (such as TXS 0506+056 \cite[][]{icecube2018,abb2024}) offer corroborative evidence of a hadronic component in the jet. 
    \item {\bf {Hybrid Jet}}: A hybrid lepton--hadronic jet has a mass composition between a pure pair-dominated/leptonic jet and a hadronic jet, encompassing the most general jet composition, which includes electrons, positrons, and baryons. Multifrequency and multi-messenger modeling efforts for the analysis of blazar observations usually include all these components ~\citep[]{Rodrigues2024}.
\end{itemize}

\subsection{Leptonic Interactions and Associated Radiative Processes} \label{sec:lep_rt}
\textls[-15]{The radiative processes associated with leptons include synchrotron, bremsstrahlung, and IC emission. The synchrotron radiation typically peaks at} a frequency of
\mbox{$\nu_{\rm syn} \sim 10^{6} \gamma^{2} B~{\rm Hz}$,}
where \(B\) is the magnetic field strength in Gauss, and \(\gamma\) is the Lorentz factor of the leptons; see, e.g., \citep[][]{dermer2009}. Accurate modeling of the spectral energy distribution (SED) requires not only the underlying radiative mechanisms but also the energy distribution of the lepton population. Rather than assuming monoenergetic leptons, it is essential to specify the number density as a function of energy for the ensemble of leptons.

In the IC process, relativistic leptons scatter incoming soft photons, possibly through single or multiple scattering events. A single scattering can boost the photon energy by a factor of up to $\sim$$\gamma^2$ relative to the frequency of seed photons \citep[][]{rybicki1986}. Synchrotron radiation is inherently highly polarized, and Compton scattering can further enhance the polarization of the upscattered high-energy photons \citep[][]{rybicki1986,dermer2009}, providing potential diagnostics for both the lepton population and the magnetic field configuration \citep[][]{ana2020}.

Bremsstrahlung emission occurs when leptons are accelerated in the Coulomb fields of ambient charged particles. Although bremsstrahlung may contribute significantly to the SED of hot and dense accretion flows ~\citep[]{mahadevan1997}, it is generally negligible in jet environments. This is due to the low particle densities in jets and the suppression of bremsstrahlung cross sections at relativistic lepton energies~\citep[]{ff_cross_2019PhRvA.100c2703J}.

\subsection{Hadronic Interactions and Associated Radiative and Particle Processes}  \label{sec:had_rt}
The hadron can interact with the photon (photo--hadronic, ${\rm p}\gamma$, process) and other hadrons (proton--proton, pp, interaction). 
The interactions of the hadronic process facilitate the formation of pions ($\pi^{0}, \pi^{+}, \pi^{-}$), whose decay subsequently results in the production of gamma rays from $\pi^{0}$ and leptons and neutrinos from $\pi^{+}$ and $\pi^{-}$.
Direct production channels related to pion production include \citep[]{dermer2009,owen2023}
\begin{equation}
    {\rm p}+\gamma\to\Delta^{+}\to\begin{cases}
    {\rm p}+\pi^{0} \\
    {\rm n} +\pi^{+}
  \end{cases}\;,
\end{equation}
and 
\begin{align}
    {\rm p}+{\rm p}\to \begin{cases}
    {\rm p}+{\rm p}+\pi^{0} \\
    {\rm p}+{\rm p} +\pi^{+}+\pi^{-}\\
    {\rm p}+{\rm n} +\pi^{+}
  \end{cases}\;.
\end{align}

{The} decay of the neutral pion produced in the above mechanism then results in strong gamma-ray emissions
\begin{align}
    \pi^{0}\to \gamma+\gamma\;,
\end{align}
and charged pions produce secondary leptons and neutrinos:
\begin{align}
    \pi^{+}\to &\mu^{+}+\nu_{\mu}\\
    &\hspace{-.2ex}\downarrow   \nonumber\\
    &{\rm e}^{+}+\nu_{\rm e}+\bar{\nu}_{\mu} \nonumber
\end{align}\vspace{-24pt}
\begin{align}
    \pi^{-}\to &\mu^{-}+\bar{\nu}_{\mu}\\
    &\hspace{-.2ex}\downarrow  \nonumber\\
    &{\rm e}^{-}+\bar{\nu}_{\rm e}+\nu_{\mu}  \nonumber
\end{align}

{The} relative importance of these hadronic processes differs across black hole mass scales.
The neutral pion production threshold in p$\gamma$ intereactions is set by the photon energy at approximately 140 MeV in the proton rest frame. In the lab frame, assuming head-on collisions, a criterion of $2\gamma_{\rm p}E_{\gamma}\ge 140$ MeV is required, which implies the need for a high-energy photon field at approximately hundreds of MeV. 
Both the p$\gamma$ and the pp channels are expected for AGN jets, within which UHECR can be produced e.g.,~\citep[]{jacobsen2015,murase2017}.
In comparison, miniquasar jets can accelerate protons less efficiently, and the gamma-ray emissions from the hadronic process would mainly be contributed by the pp process.

\section{EMRB  as a Probe to Jet Composition}\label{S3}
The periodic interactions between the accretion flow of the primary SMBH and the secondary miniquasar present a distinctive opportunity to examine the composition of the miniquasar jet. As illustrated in phase (b) of Figure \ref{fig:overview}, the collision between the miniquasar jet and the accretion flow of the SMBH can lead to significant variations in emission characteristics, with specific details associated with the jet's composition. For instance, as opposed to a pair-dominated jet, the collision can result in proton--proton interactions within a baryonic jet.

In the following, after systematically investigating the parameter space for  EMRBs of interest,  we model the orbital dynamics and the UV and gamma-ray light curves dependent on orbital movement for a representative scenario, and perform a comparative analysis of the emission signatures that correspond to different jet composition.

\subsection{Orbital Parameter Space}\label{sec:orbital}

\begin{figure}[H]
\includegraphics[width=0.8\textwidth]{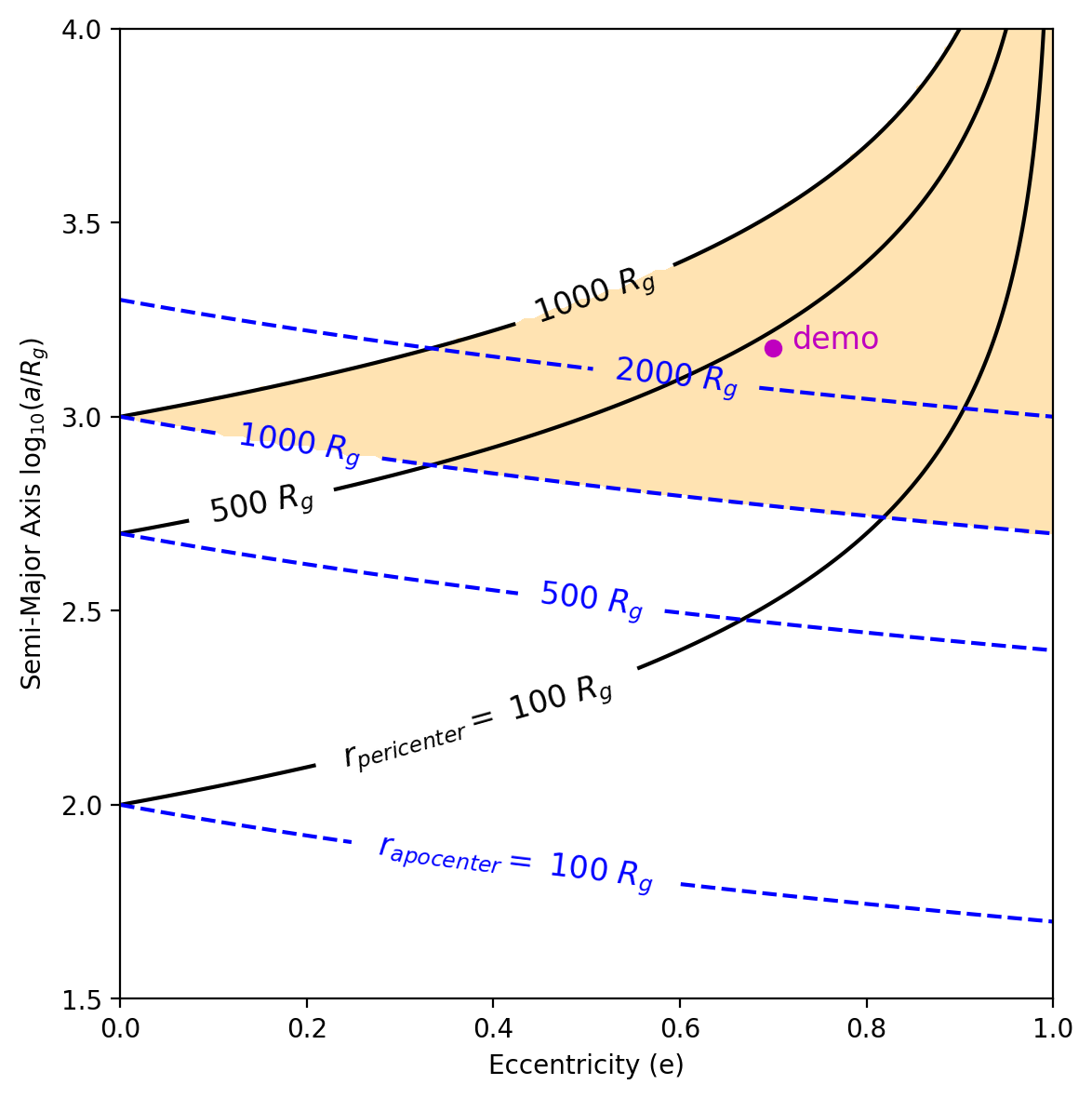}
\caption{{Pericenter} 
 and apocenter radii as functions of the semi-major axis $a$ and the eccentricity $e$. miniquasars with orbital parameters falling within the shaded region have pericenters smaller than $10^{3}~R_{\rm{g}}$ and apocenters exceeding $10^{3}~R_{\rm{g}}$. Such orbits are expected to produce repeated interactions with the SMBH accretion disk, consistent with the configuration illustrated in Figure~\ref{fig:overview}. The selected parameters for our demonstrative case is indicated by the magenta dot.
}\label{fig:pericenter}
\end{figure}

In addition to the semi-major axis $a$  (see  Figure \ref{fig:period}), the orbital characteristics of the miniquasar are also governed by its eccentricity $e$. Figure \ref{fig:pericenter} presents the pericenter \mbox{$r_{\rm peri}=a(1-e)$} and apocenter \mbox{$r_{\rm apo}=a(1+e)$} within the parameter space. As demonstrated in Figure \ref{fig:overview}, our focus is on  EMRBs  within which the miniquasar undergoes repeated interactions with SMBH accretion by entering and leaving the SMBH accretion disk. With an outer boundary of the SMBH disk of $\sim$$10^3 R_{g}$, the shaded region in Figure \ref{fig:overview} satisfies these criteria.

With a given  semi-major axis $a$, eccentricity $e$, and the mass of SMBH $M_{\rm SMBH}$, the orbit of the miniquasar can be computed by  Kepler’s equation \citep{Odell1986,murray1999}, 
which relates the mean anomaly $M(t)$ to the eccentric anomaly $E(t)$ via
\begin{align}
M(t) &= E(t) - e \sin E(t)\;.
\end{align}

{The} mean anomaly evolves linearly in time:
\begin{align}
M(t) &= \frac{2\pi}{T}(t - t_0)\;,
\end{align}
with $T$ being the orbital period and $t_0 $ the time of pericenter passage, which is set to $t_0 = 0$ without loss of generality. Once $E(t)$ is determined numerically, the radius to the SMBH can be recovered from
\begin{align}\label{eq:r(t)}
r(t) &= a(1 - e \cos E(t)).
\end{align}

{Together,} these equations provide a complete parametric description of orbital motion, linking geometrical shape with temporal evolution.

\subsection{Modeling Multi-Frequency Emissions of the Miniquasar Jet}\label{sec:modeling_emisssion}
The emission of jets at different frequencies can be quantified by evaluating the transformation of jet power into emission power, guided by the radiation mechanisms described in  Sections \ref{sec:lep_rt} and \ref{sec:had_rt}. We focus on the energetic event during the miniquasar jet's interaction with the SMBH accretion (phase (b) in Figure \ref{fig:overview}). In such cases, the magnetic field strength and the photon field are set by the local conditions 
 of the accretion disk. 

For a miniquasar jet interacting with the thin disk of a SMBH, without losing generality, we consider a bulk Lorentz factor of
$\Gamma\simeq5$ cf. \citep[][]{Fender2004}.  The 
average particle Lorentz factors have values
in the range  $\langle\gamma_{\rm e}\rangle\sim10^{2-3}$ and
$\langle\gamma_{\rm p}\rangle\sim10^{0-1}$ (cold protons). 
We estimate the lepton emissions as follows (see also Section \ref{sec:lep_radiation}).
For leptons with a Lorentz factor 
$\gamma  \sim\Gamma\langle\gamma_{\rm e}\rangle\sim10^{3}$ and 
a magnetic field strength of 
  $B\sim 10^{2-3}\;\! {\rm G}$ in the accretion disk,  
  the synchrotron frequency $\nu_{\rm syn}\sim 10^{6}\gamma^{2}B(\rm Gauss)$ would be  $\sim$$10^{14-17}$ Hz , 
  which is in the optical/ultraviolet (UV) bands  
  in the co-moving frame moving of the accretion flow. 
The UV radiation emitted by the disk can be enhanced through IC scattering. As a rough estimate, the frequency of Compton-scattered radiation (with a single scattering event) is $\nu_{\rm IC}\sim 10^{20-24}~{\rm Hz}$. This frequency corresponds to energies in the MeV/GeV range.

For the hadronic process, the energy of the gamma rays from the decay of $\pi^0$ is around $0.1E_{\rm p}$, 
where $E_{\rm p}\sim\Gamma\langle\gamma_{\rm p}\rangle m_{\rm p}c^2$.  
The output gamma-ray also contributes to the GeV photonic radiation. 
Secondary leptons from the decay of charged pion possess an energy similar to that of $0.1E_{\rm p}$, corresponding to $\langle\gamma_{\rm e}^{\rm secondary}\rangle\sim 10^{3}$, which is thus similar to that of the \mbox{primary leptons.}  

With the above-mentioned potential output radiation ranges in mind, the relative power between optical/UV  and MeV/GeV  emissions indicates the relative amounts of leptons and hadrons in the jet. 
For the purpose of examining the general multi-frequency emission characteristics,
we do not further distinguish between the optical/UV and MeV/GeV bands in the subsequent discussion, and treat the synchrotron emissions as UV radiation, and both the IC and neutral pion decay emissions as GeV gamma-ray radiation. 
A comprehensive SED modeling, which requires detailed input energy spectra of the relevant particles and photon fields, is deferred to future work.

The related notations for the later estimation of UV and gamma-ray power are summarized in Table \ref{tab:notation}: the electromagnetic properties of the jet emissions from miniquasars are influenced by three main factors. These are the jet composition, symbolized by $\xi_{(\cdots)}$ (see Equation~(\ref{eq:xi})); the portion of power attributed to different types of particles, represented by $\eta_{(\cdots)}$ (see Equation~(\ref{eq:eta_e})); and the efficiency of the conversion of particle energy, indicated by $f_{(\cdots)}$. The subscript $_{(\cdots)}$ specifies the associated species or radiative processes.

\begin{table}[H]
\centering
\caption{Summary of notations related to composition of jet content, jet power, and radiation power.}\label{tab:notation}
\begin{adjustwidth}{-\extralength}{0cm}
\begin{tabular}{l m{11.3cm}l}
\toprule
\textbf{Notation} & \textbf{Definition} & \textbf{Reference}\\
\midrule
$\xi_{\rm e^{+}}$ & positron to electron ratio& Equation~(\ref{eq:xi})\\
\addlinespace
$\xi_{\rm p}$ & proton to electron ratio  & Equation~(\ref{eq:xi})\\
\addlinespace
$\eta_{\rm e^{-}}$ & jet power carried by electron &Equation~(\ref{eq:eta_e})\\
\addlinespace
$\eta_{\rm e^{+}}$ & jet power carried by positron &Equation~(\ref{eq:eta_e})\\
\addlinespace
$\eta_{\rm p}$ & jet power carried by proton &Equation~(\ref{eq:eta_e})\\
\addlinespace
$f_{\rm IC}$ & lepton power to  $\gamma$-ray power (via IC process) conversion rate &Equations~(\ref{eq:lepton_to_gamma}) and~(\ref{eq:2lepton_to_gamma})\\
\addlinespace
$f_{\rm syn}$ & lepton power to UV  power (via synchrotron process) conversion rate &Equations~(\ref{eq:lepton_to_UV}) and~(\ref{eq:2lepton_to_syn})\\
\addlinespace
$f_{pp}$ & efficiency of pp interaction & Equation~(\ref{eq:fpp=1})\\
\addlinespace
$f_{\pi^{0}\to\gamma{\rm-ray}}$ & neutral pion power to  $\gamma$-ray power  conversion rate & Equation~(\ref{eq:pion0decay})\\
\addlinespace
$f_{\pi^{\pm}\to e^{\pm}\to\gamma{\rm-ray}}$ & charged pion power to $\gamma$-ray power (via IC process of secondary leptons) conversion rate & Equation~(\ref{eq:2lepton_to_gamma})\\
\addlinespace
$f_{\pi^{\pm}\to e^{\pm}\to\rm UV}$ & charged pion power to UV power (via synchrotron process of secondary leptons) conversion rate & Equation~(\ref{eq:2lepton_to_syn})\\

\bottomrule
\end{tabular}\\
\end{adjustwidth}
\label{tab:parameters}
\end{table}

\subsubsection{Leptonic Emission}\label{sec:lep_radiation}
The gamma-ray emissions contributed from the lepton composition (both electron and positron) of the jet are generally associated with the IC process. The estimated gamma-ray power is
\begin{align}\label{eq:lepton_to_gamma}
    P_{{\rm e}^{\pm}\to {\gamma{\rm-ray}}}\approx  f_{\rm IC} (P_{\rm e^{-}}+P_{\rm e^{+}}) =f_{\rm IC}(1+\xi_{{\rm e}^{+}})\eta_{{\rm e}^{-}}P_{\rm jet}\;.
\end{align}

{The} energy conversion rate, denoted as $f_{\rm IC}$, refers to the process by which electrons and positrons transfer their energy to photons through the IC process.
While the leptonic energy can be converted into scattered, gamma-ray photons, part of the energy can also be converted to synchrotron, UV photons, with the power 
\begin{align}\label{eq:lepton_to_UV}
    P_{{\rm e}^{\pm}\to {\rm UV}}\approx f_{\rm syn}(1+\xi_{{\rm e}^{+}})\eta_{{\rm e}^{-}}P_{\rm jet}\;.
\end{align}

{The} ratio between UV and gamma-ray power can be estimated as the ratio between the energy of the ambient radiation field $U_{\rm ph}$ and the local magnetic field energy $U_{\rm B}$:
\begin{align}\label{eq:y_parameter}
    \dfrac{P_{{\rm e}^{\pm}\to {\gamma{\rm-ray}}}}{P_{{\rm e}^{\pm}\to {\rm UV}}}=\dfrac{f_{\rm IC}}{f_{\rm syn}}\approx\frac{U_{\rm ph}}{U_{\rm B}}\;.
\end{align}

{The} conversion of lepton energy to emissions via the IC or synchrotron processes is not entirely efficient. Assuming that the sum of the energy conversion from the processes is approximately constant, as denoted by $\mathcal{Q}$, we impose the following constraint:
\begin{align}\label{eq:constraint_fIC}
f_{\mathrm{IC}} + f_{\mathrm{syn}} = \mathcal{Q} < 1 \;.
\end{align}

{This} condition reflects the fact that only a fraction of the available particle energy is radiated away via synchrotron and IC processes, with the remainder either being retained by particles or lost through non-radiative channels such as adiabatic expansion or escape.

\subsubsection{Hadronic Emission}\label{sec:had_radiation}
The gamma-ray emissions that emanate from the baryonic jet are not only derived from primary electrons but also arise through hadronic processes, such as pp or p$\gamma$ interactions. The interactions of the hadronic process facilitate the formation of pions $\pi^{0}, \pi^{+}, \pi^{-}$, whose decay subsequently results in the production of gamma rays from $\pi^{0}$ and electrons/positrons from $\pi^{+}, \pi^{-}$. As discussed in Section \ref{sec:had_rt}, the pp interaction between the miniquasar jet and the SMBH accretion flow emerges as a more plausible mechanism for the production of pion. The produced gamma-ray power from $\pi^{0}$ decay (with energy conversion rate $f_{\pi^{0}\to\gamma{\rm-ray}}$) and secondary leptons (with energy conversion rate $f_{\pi^{\pm}\to e^{\pm}\to\gamma{\rm-ray}}$) can be estimated by 
\begin{equation}\label{eq:hadron_to_gamma_GeV}
\begin{array}{ll}
    P_{{\rm p}\to \gamma{\rm-ray}}&\approx f_{\rm pp}(\dfrac{1}{3}f_{\pi^{0}\to\gamma{\rm-ray}}+\dfrac{2}{3}f_{\pi^{\pm}\to e^{\pm}\to\gamma{\rm-ray}})P_{\rm p}\\
    &=f_{\rm pp}(\dfrac{1}{3}f_{\pi^{0}\to\gamma{\rm-ray}}+\dfrac{2}{3}f_{\pi^{\pm}\to e^{\pm}\to\gamma{\rm-ray}})\dfrac{{\rm m}_{\rm p}\langle\gamma_{\rm p}\rangle}{{\rm m}_{\rm e}\langle\gamma_{\rm e}\rangle}\xi_{\rm p}\eta_{\rm e^{-}}P_{\rm jet}\;,
\end{array}
\end{equation}
where $f_{\rm pp}$ is the efficiency of the pp interactions.
The coefficients $1/3$ and $2/3$ in \mbox{Equaiton (\ref{eq:hadron_to_gamma_GeV})}  are estimated by the cross section of the pion production in the pp collision \citep[][]{dermer2009,owen2023}. 
The secondary leptons from pion decay can also contribute to synchrotron emissions at the UV band:
\begin{equation}\label{eq:hadron_to_UV}
    P_{{\rm p}\to {\rm UV}}\approx f_{\rm pp}(\dfrac{2}{3}f_{\pi^{\pm}\to {\rm e}^{\pm}\to\rm UV})\dfrac{{\rm m}_{\rm p}\langle\gamma_{\rm p}\rangle}{{\rm m}_{\rm e}\langle\gamma_{\rm e}\rangle}\xi_{\rm p}\eta_{\rm e^{-}}P_{\rm jet}\;,
\end{equation}
where $f_{\pi^{\pm}\to {\rm e}^{\pm}\to\rm UV}$ represents the energy conversion rate to UV emissions. The various energy conversion rates,  $f_{(\cdots)}$, are estimated in the following. 

To approximate $f_{\rm pp}$, one can examine the correlation between the dynamical timescale $t_{\rm dyn}$ and the interaction timescale $t_{\rm pp}$:
\begin{equation}\label{eq:fpp}
 f_{\rm pp}={\rm min}(1, \dfrac{t_{\rm dyn}}{t_{\rm pp}})={\rm min}(1, \dfrac{H/(\Gamma c)}{1/(n_{\rm p}\sigma_{\rm pp}c)})\;. 
\end{equation}

{The} \textls[-25]{estimated dynamical timescale is roughly $H/(\Gamma c)\sim (0.01~R)/(\Gamma c)\sim10^{2}~{\rm s}$, assuming that a jet with $\Gamma=5$ passes through the SMBH accretion disk at $R\sim100~ GM_{\rm SMBH}/c^{2}\sim 10^{15}~{\rm cm}$} for $M_{\rm SMBH}=10^{8}~{{\rm M}_{\odot}}$, and that the disk height is $H\sim 0.01~R$. The estimated interaction timescale is $t_{\rm pp}<1~{\rm s}$, by adopting the cross section $\sigma_{\rm pp}\sim30$ mb ($1~{\rm mb}=10^{-27}$ cm$^{-2}$) \citep[][]{kafexhiu2014}, and $n_{\rm p}\sim 10^{16-18}~{\rm cm}^{-3}$ (as estimated by the number density of the thin disk solution for the primary SMBH  
\cite[][]{Shakura1973,nov73}). Therefore, we estimate
\begin{align}\label{eq:fpp=1}
    f_{\rm pp}\approx 1\;.
\end{align}

{In} addition, the decay of neutral pions into gamma rays is highly efficient:
\begin{align}\label{eq:pion0decay}
f_{\pi^{0} \to \gamma\text{-ray}} \approx 1\;.
\end{align}

{Equations} (\ref{eq:fpp=1}) and (\ref{eq:pion0decay}) suggest that the interaction between the jet and disk allows for a highly efficient transformation of the proton power into gamma-ray emissions, highlighting a key characteristic of the EMRBs  under study.

\textls[-15]{In comparison, the efficiency of converting charged pion energy to radiation includes the transformation of charged pion energy into lepton energy, denoted} as \mbox{$f_{\pi^{\pm}\to {\rm e}^{\pm}}(\approx$$0.25 $ \citep{kelner2006,owen2023}),} followed by the conversion of lepton energy into IC or synchrotron radiation energy. We estimate the following efficiency:
\begin{align}\label{eq:2lepton_to_gamma}
    f_{\pi^{\pm}\to {\rm e}^{\pm}\to\gamma{\rm-ray}}=f_{\pi^{\pm}\to {\rm e}^{\pm}}~f_{\rm IC}\approx 0.25f_{\rm IC}
\end{align}
and
\begin{align}\label{eq:2lepton_to_syn}
    f_{\pi^{\pm}\to {\rm e}^{\pm}\to\rm UV}=f_{\pi^{\pm}\to {\rm e}^{\pm}}~f_{\rm syn}\approx 0.25f_{\rm syn}\;.
\end{align}

{The} secondary leptons therefore follow a similar relation as in Equation (\ref{eq:y_parameter}):

\begin{align}\label{eq:y_parameter_pion}
    \frac{f_{\pi^{\pm}\to {\rm e}^{\pm}\to\gamma{\rm-ray}}}{f_{\pi^{\pm}\to {\rm e}^{\pm}\to\rm UV}}=\dfrac{f_{\rm IC}}{f_{\rm syn}}\approx\frac{U_{\rm ph}}{U_{\rm B}}\;.
\end{align}

{It} is assumed that the constraint for $f_{\rm IC}$ and $f_{\rm syn}$, Equation (\ref{eq:constraint_fIC}), also applies to the secondary leptons, and therefore 
\begin{align}\label{eq:constraint_fpion}
    f_{\pi^{\pm}\to {\rm e}^{\pm}\to\gamma{\rm-ray}}+f_{\pi^{\pm}\to {\rm e}^{\pm}\to\rm UV}\approx0.25\mathcal{Q}\;.
\end{align}

\subsubsection{Estimation of Emission Efficiency and Multi-Frequency Emission Ratio}\label{sec:ratio_estimation}
 With the conversion rates from the particle energy to the radiation, $f_{(\cdots)}$, we can estimate the radiation power of jets with a different composition. 
The multi-wavelength radiation power from the jet power carried by the leptonic and hadronic components\linebreak   is, respectively,
\begin{align}
    \dfrac{P_{{\rm e}^{-}\to\gamma{\rm-ray}}+P_{{\rm e}^{-}\to{\rm UV}}}{P_{{\rm e}^{-}}}=\dfrac{P_{{\rm e}^{+}\to\gamma{\rm-ray}}+P_{{\rm e}^{+}\to{\rm UV}}}{P_{{\rm e}^{+}}}=f_{\rm IC}+f_{\rm syn}=\mathcal{Q}<1\;,
\end{align}
and
\begin{adjustwidth}{-\extralength}{0cm}
\begin{equation}
    \frac{P_{{\rm p} \to \gamma\text{-ray}} + P_{{\rm p} \to \text{UV}}}{P_{{\rm p}}}
    \approx f_{pp} \left[
        \frac{1}{3} f_{\pi^{0} \to \gamma\text{-ray}} + \frac{2}{3} \left(
            f_{\pi^{\pm} \to e^{\pm} \to \gamma\text{-ray}} + f_{\pi^{\pm} \to e^{\pm} \to \text{UV}}
        \right)
    \right] = \left( \frac{1}{3} + \frac{\mathcal{Q}}{6} \right) < 1 \;.
\end{equation}
\end{adjustwidth}

{The} ratios are constant if $\mathcal{Q}$ is constant over time.

\begin{figure}[H]
\includegraphics[width=0.9\textwidth]{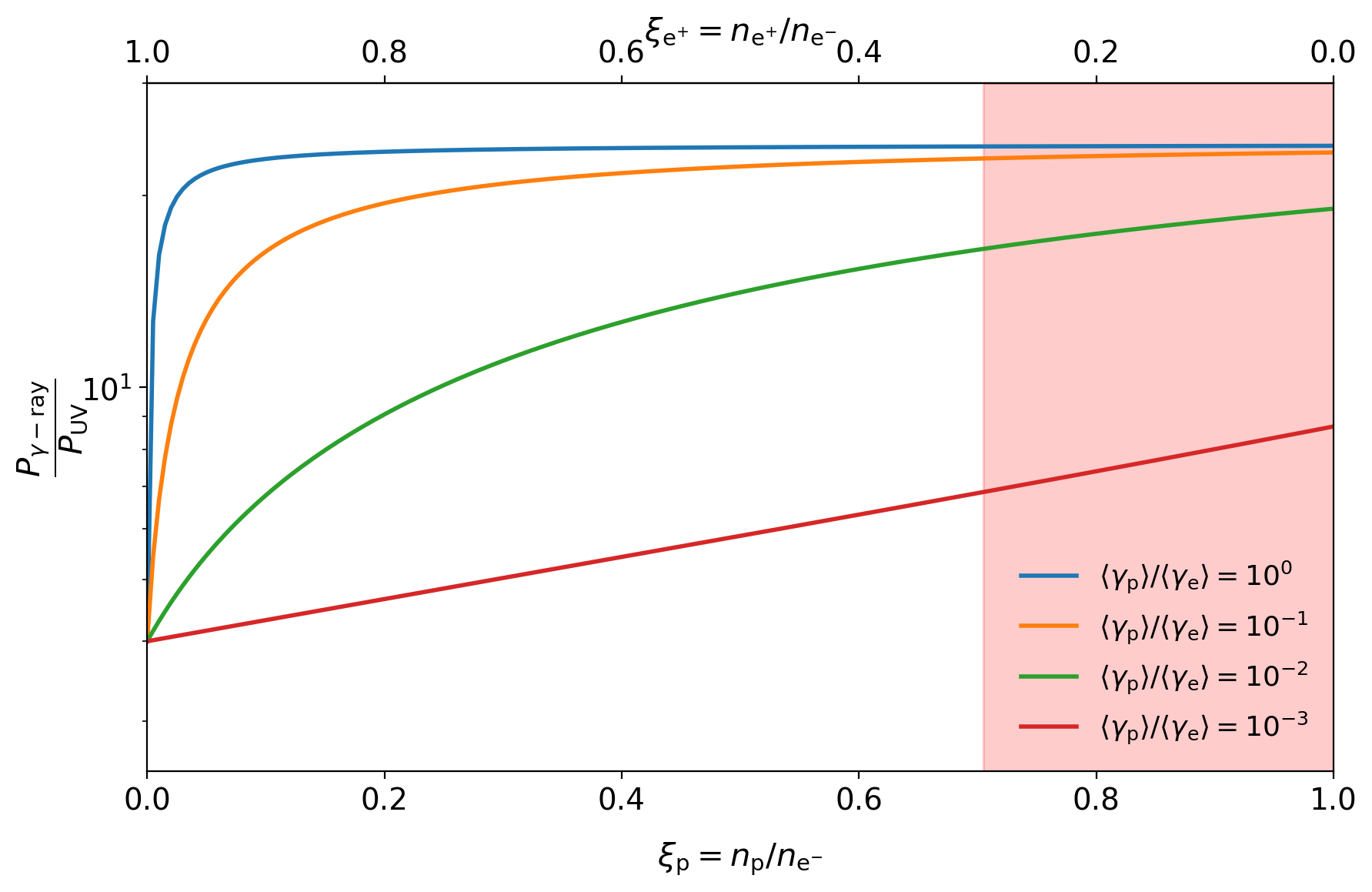}
\caption{{Estimated} 
 gamma-ray-to-UV power ratio for varying ratios between the particle Lorentz factor of proton and leptons, $\langle\gamma_{\rm p}\rangle/\langle\gamma_{\rm e}\rangle$; see Equation (\ref{eq:MeV_to_baryonic}). The parameters $\mathcal{Q}=0.5$  and $f_{\rm IC}=0.8\mathcal{Q}$ are assumed. The shaded region corresponds to cases in which the baryonic component accounts for more than 99.9$\%$ of the total mass.\label{fig:ratios_compare}}
\end{figure}

For a hybrid jet, the estimated gamma-ray-to-UV emission ratio  is
\begin{align}\label{eq:MeV_to_baryonic}
    \dfrac{P_{\gamma{\rm-ray}}}{P_{\rm UV}}=\dfrac{P_{{\rm p}\to\gamma{\rm-ray}}+{P_{{\rm e}^{\pm}\to\gamma{\rm-ray}}}}{P_{{\rm p}\to{\rm UV}}+P_{{\rm e}^{\pm}\to{\rm UV}}}\approx\dfrac{(\dfrac{1}{3}+\dfrac{f_{\rm IC}}{6})\dfrac{m_{\rm p}\langle\gamma_{\rm p}\rangle }{m_{\rm e}\langle\gamma_{\rm e}\rangle} \xi_{\rm p}+(1+\xi_{\rm e^{+}})f_{\rm IC}}{\dfrac{\mathcal{Q}-f_{\rm IC}}{6}\dfrac{m_{\rm p}\langle\gamma_{\rm p}\rangle }{m_{\rm e}\langle\gamma_{\rm e}\rangle} \xi_{\rm p}+(1+\xi_{\rm e^{+}})(\mathcal{Q}-f_{\rm IC})}\;,
\end{align}
which is related to the mass content, $\xi_{\rm p}$ and $\xi_{\rm e^{+}}~(=1-\xi_{\rm p})$, as well as the energy content, $\langle\gamma_{\rm p}\rangle/\langle\gamma_{\rm e}\rangle$, within the jet.
{In Figure} \ref{fig:ratios_compare} we present the ratio of gamma-ray-to-UV power with different mass and energy content, assuming $\mathcal{Q}=0.5$ (50$\%$ leptonic power within the jet goes to radiation power) and $f_{\rm IC}=0.8\mathcal{Q}$ (the IC process dominates the leptonic radiation).  
 When $\xi_{\rm p}\to 0$, the jet is pair-dominated and all cases reduced to
\begin{align}\label{eq:MeV_to_UV_pair}
    \dfrac{P_{{\rm e}^{\pm}\to {\gamma{\rm-ray}}}}{P_{{\rm e}^{\pm}\to {\rm UV}}}=\dfrac{f_{\rm IC}}{f_{\rm syn}}\approx\dfrac{f_{\rm IC}}{\mathcal{Q}-f_{\rm IC}}\;,
\end{align}
which has a value $f_{\rm IC}/(\mathcal{Q}-f_{\rm IC})=4$ for this setup.
As the ratio $\langle\gamma_{\rm p}\rangle/\langle\gamma_{\rm e}\rangle$ increases, protons account for a greater portion of the jet's power (refer to Figure \ref{fig:eta_compare}), which also results in a higher gamma-ray-to-UV power ratio. This is due to the effective gamma-ray production by pp collisions.

\subsection{Two-Stage Accretion and Jet Power }\label{sec:two-stage}
As the miniquasar undergoes recurrent orbital phases, episodic accretion and jet-launching are expected to occur when it passes through the dense region of the SMBH accretion flow.  An estimate of the passing time for the miniquasar with orbital velocity $v_{\rm orb}$ to pass through the SMBH disk with disk height $H$, $t_{\rm pass}=H/v_{\rm orb}$, is multiple hours
\endnote{
Assuming the miniquasar passes the SMBH disk around the pericenter $r_{\rm peri}$, we have $H=0.01~r_{\rm peri}$ and $v^{2}_{\rm orb}=GM_{\rm SMBH}/r_{\rm peri}$, 
\begin{equation*}
    t_{\rm pass}=H/v_{\rm orb}\sim  5\times10^{-8} (\frac{r_{\rm peri}}{R_{\rm g}})^{3/2} (\frac{M_{\rm SMBH}}{M_{\odot}}) ~{\rm s}.
\end{equation*}
{In} 
 order of magnitude $r_{\rm peri}\sim10^2~R_{\rm g}$, and $M_{\rm SMBH}=10^{8}~M_{\odot}$, $t_{\rm pass}$ is about a few hours.}. Assuming the viscous time scale $t_{\rm visc}$ is much longer than the passing time (e.g., months), the accretion of the miniquasar can be considered using a two-stage process.
The first stage is mass capture through Bondi--Hoyle accretion \citep[][]{bon44,bon52} while it crosses the SMBH accretion flow  with a supersonic velocity. The captured mass $\Delta M_{\text{cap}}$ is approximately given by
\begin{equation}\label{eq:m_cap}
\Delta M_{\text{cap}} \approx \rho_{\text{disk}} \times \pi \left( \frac{2 G M_{\text{mq}}}{v_{\text{rel}}^2} \right)^2 \times v_{\text{rel}} \times t_{\text{pass}}
\end{equation}
where
$\rho_{\text{disk}}$ is the local gas density, 
$M_{\text{mq}}$ is the miniquasar mass, 
$v_{\text{rel}}\sim v_{\rm orb}$ is the relative velocity between the miniquasar and the disk gas, and
$t_{\text{pass}}$ is the disk crossing timescale.  The captured mass is related to $\rho_{\rm disk}$, $M_{\rm mq}$, and $H(\simeq$$v_{\rm rel} \times t_{\rm pass})$. The capture radius, $2 G M_{\text{mq}}/v_{\text{rel}}^2$, is roughly ten times smaller than the disk height. In general, for a fixed mass and accretion rate of SMBH, $\Delta M_{\rm cap}$ increases with increasing pericenter radius due to the combined effects of orbital velocity, disk density, and disk height as functions of radius (see Appendix \ref{app:Mcap_mq} for estimates of $\Delta M_{\rm cap}$).

The second stage is that in which the miniquasar accretes the captured materials  until all available accreting material is exhausted. The evolution of the accretion rate is modeled as a viscous decay \citep{pringle1981,tanaka2011}:
\begin{align}\label{eq:mdot_t}
    \dot{M}_{\rm mq}(t) =\dot{M}_{\rm mq}(t=0) \times \left( 1 + \frac{t}{t_{\text{visc}}} \right)^{-\gamma}\;,
\end{align}
 with normalization 
\begin{equation}\label{eq:mdot_norm}
    \dot{M}_{\rm mq}(t=0)=\frac{\Delta M_{\rm cap}}{2~t_{\rm visc}}\;,
\end{equation}
where the viscous timescale is $t_{\text{visc}}$ and $\gamma=3/2$ is adopted for demonstrative purposes. Here, we treat $t_{\text{visc}}$ as a free parameter\endnote{
A typical estimate of the viscous timescale is given by ~\citep[]{kato1998}
\[
t_{\rm visc} \sim \frac{R_{\rm mq}^2}{\nu},
\]
where \(R_{\rm mq}\) is the initial radius of the accreting material {{around} 
 the miniquasar}, \( \nu \) is the kinematic viscosity.  
Using the standard prescription \(\nu \sim \alpha c_{\rm s} H_{\rm mq}\) with \( H_{\rm mq} \sim c_{\rm s} / \Omega_{\rm k, mq} \), where \( \alpha \) is the dimensionless viscosity parameter, \( c_{\rm s} \) is the sound speed {{within} the disk}, {\( H_{\rm mq} \) is the vertical scale height of the miniquasar disk  , and \( \Omega_{\rm k,mq} \) is the Keplerian angular velocity}, we obtain
\[
t_{\rm visc} \sim \frac{1}{\alpha} \left( \frac{R_{\rm mq}}{H_{\rm mq}} \right)^{2} \frac{1}{\Omega_{\rm k,mq}}.
\]

{For} a  slim-disk case, \( R_{\rm mq}/H_{\rm mq} \sim 1-10 \), this yields
\[
t_{\rm visc} \sim 50 (\dfrac{\alpha}{0.1})^{-1}\left( \frac{R_{\rm mq}}{H_{\rm mq}} \right)^{2}(\frac{R_{\rm mq}}{100R_{\rm g,mq}})^{3/2}( \frac{M_{\rm mq}}{ 10^{3}M_{\odot}}) \ \mathrm{s},
\]
{{where} $R_{g,mq}\equiv { GM_{\rm mq}}/c^{2}$}.
 The time scale given by the above estimation {{with} a Bondi radius of order $R_{\rm mq}\sim 10^{2} R_{\rm g,mq}$} seems too short. In this work, we instead treat \( t_{\rm visc} \) as a free parameter, empirically constrained by the jet-launching condition provided in Equation~(\ref{eq:mdot_constraint1}).
}.  Assuming $t_{\rm visc}$ is of the order of months, the resulting accretion for the miniquasar is super-Eddington. That is, $ \dot{M}_{\rm mq} / \dot{M}_{\rm Edd} >1$, with the Eddington accretion rate $ \dot{M}_{\rm Edd} \equiv L_{\rm Edd}/(0.1c^2) $  and $ L_{\rm Edd}\sim 10^{38} ({M_{\rm BH}}/{\rm M}_{\odot})~\mathrm{erg\,s^{-1}} $ is the Eddington luminosity, as demonstrated in Appendix \ref{app:Mcap_mq}.  In our demonstrative example, we assume the mass of the miniquasar is $M_{\rm mq}=500~M_{\odot}$ (see also Figure \ref{fig:Mcap_500} and Table \ref{tab:parameters}).

\begin{table}[H]
\caption{{Orbital} 
 and emission parameters of the   demonstrative EMRB system.}  
\tabcolsep=0.4cm
\begin{tabular}{ccc}
\toprule
\textbf{Orbital-Related Parameters} & \textbf{Value} & \textbf{Reference}\\
\midrule
\addlinespace
$M_{\rm SMBH}$&$10^{8}~{\rm M}_{\odot}$& Section~\ref{sec:orbital}\\
\addlinespace
$a$&1500 $R_{g}$& Section~\ref{sec:orbital}\\
\addlinespace
$e$&0.7& Section~\ref{sec:orbital}\\
\midrule
emissions-related parameters & value & note\\
\midrule
\addlinespace
$\langle\gamma_{\rm p}\rangle/\langle\gamma_{\rm e}\rangle$&$10^{-2}$&Equation (\ref{eq:eta_e})\\
\addlinespace
$\mathcal{Q}$&0.5&Equations (\ref{eq:constraint_fIC}) and  (\ref{eq:constraint_fpion})\\
\addlinespace
$f_{\rm IC}$&0.8~$\mathcal{Q}$& Equations~(\ref{eq:y_parameter}) and (\ref{eq:y_parameter_pion}) \\
\addlinespace
$f_{\rm pp}$&1&Equation (\ref{eq:fpp=1})\\
\addlinespace
$f_{\pi^{0}\to \gamma{\rm-ray}}$&1&Equation (\ref{eq:pion0decay})\\
\midrule
miniquasar accretion parameters & value & note\\
\midrule
$M_{\rm mq}$  & 500 $M_{\odot}$&related to Equation (\ref{eq:m_cap})\\
$t_{\rm visc}$& 2 weeks& related to Equations (\ref{eq:mdot_t}) and (\ref{eq:mdot_norm})\\
\bottomrule
\end{tabular}\\
\label{tab:parameters}
\end{table}

We are especially interested in jet-launching together with transient accretion.  The miniquasar accretion flow with $ \dot{M}_{\rm mq} / \dot{M}_{\rm Edd} >1$ is of the slim disk type, e.g.,~\citep[]{slim1988, kato1998}, and radiation pressure dominated jet can be launched \citep[]{skadowski2014,ricarte2023,curd2023}. Applying such empirical disk–jet coupling to the miniquasar provides a useful constraint on the jet power and on the viscous timescale \( t_{\rm visc} \) so that the miniquasar jet can be triggered  or quenched. In order to maintain the miniquasar accretion within a regime that supports jet activity,  it is required that\endnote{At lower accretion rates,
observations of both black hole X-ray binaries (BHXBs) and AGN indicate that relativistic jets can be launched during a spectral transition near the accretion rate of about \mbox{\( \dot{M} / \dot{M}_{\rm Edd} = \dot{M}_{\rm crit}\approx 0.01-0.1 \) \citep[][]{esin1997,Fender2004,remillard2006,trump2011},} and the jet is quenched beyond the threshold value $\dot{M}_{\rm crit}$. As the Super-Eddington accretion continues evolves and the accretion rate decays, the relativistic jet may be quenched with the condition in Equation (\ref{eq:mdot_constraint1}) is satisfied, then launched again near the threshold value $\dot{M}_{\rm crit}$. As a result, EMRB systems may also provide opportunities to probe the jet composition for jets launched at different accretion rates.
}
\begin{equation}\label{eq:mdot_constraint1}
     \dfrac{\dot{M}_{\rm mq}}{ \dot{M}_{\rm Edd}}\ge1\;.
\end{equation} 

{Assuming} a typical super-Eddington accretion with $\dot{M_{\rm mq}} / \dot{M}_{\rm Edd}\sim 10$, a typical miniquasar intrinsic jet power $P_{\rm jet}$ can be estimated  by 
\begin{align}\label{eq:pjet}
    P_{\rm jet}=\eta\dot{M}_{\rm mq}c^{2}\;.
\end{align} 

{With} $\eta=0.1$, 
\( P_{\rm jet} \sim \dot{M}_{\rm crit} \times L_{\rm Edd} \sim 10^{42\text{--}44}~\mathrm{erg\,s^{-1}} \), with the range of miniquasar mass \( M_{\rm mq} = 10^{3\text{--}5} ~{\rm M}_{\odot} \). Potential Doppler effects due to the bulk motion of the jet may further enhance the observed jet power  (Equation (\ref{eq:Pjet_obs})).

{As} 
 a demonstration case, we consider a  EMRB system with the following parameters: $a = 1500~R_{g}$,  $e = 0.7$,   $M_{\rm SMBH} = 10^{8}~{\rm M}_{\odot}$ (corresponding to the magenta points in \mbox{Figures~\ref{fig:period} and~\ref{fig:pericenter}),} $M_{\rm mq}=500~M_{\odot}$ and $t_{\rm visc}=$ two weeks (see the magenta point in \mbox{Figure \ref{fig:Mcap_500}).}
These parameters are summarized in Table \ref{tab:parameters}. 
The orbital motion,  \mbox{Equation (\ref{eq:r(t)}),} is computed and presented in the panel (a) of \mbox{Figure \ref{fig:lc}.}
The orbital phase-dependent  accretion rate \mbox{(Equation (\ref{eq:mdot_t})) } is shown in panel (b). The assumption is made that the miniquasar jet is initiated and launched near the pericenter, $t \approx 0$.  This choice  of such $t_{\rm visc}$ is consistent with the  jet-launching conditions provided in Equation (\ref{eq:mdot_constraint1}) (see also Figure \ref{fig:Mcap_500}). { In the panel (b) of Figure \ref{fig:lc}}, the gray regions within the panel demonstrate instances where the jet-launching condition, as specified by Equation (\ref{eq:mdot_constraint1}), is not satisfied. Consequently, jet activity is expected to stop.
As illustrated, the resulting jet lifetime is shorter than the orbital period. 

In the computations, for demonstrative purpose, the jet direction is presumed to remain constant within the inertial frame, thus ensuring that the Doppler boosting effect due to the jet's bulk motion remains invariant with orbital time, and the jet power is simply proportional to Equation (\ref{eq:pjet}),  provided that the jet is launched. In a more realistic and sophisticated scenario, the jet's orientation may change with respect to the orbital phase, resulting in an orbital-dependent Doppler factor, as described by Equation (\ref{eq:doppler_factor}). The variation in the direction of the jet would also be correlated with the angle and column density at the points where the jet impacts and penetrates the disk. 

\subsection{Multi-Frequency Light Curves as Diagnostic for Jet Composition}\label{sec:lc}
Next, we model the time-dependent emission features during the collision of the miniquasar jet with the SMBH accretion flow. The multi-wavelength light curves are related to both the total jet power and the detailed emissions related to the leptonic and hadronic components within the jet.

\begin{figure}[H]
\includegraphics[width=1.\textwidth]{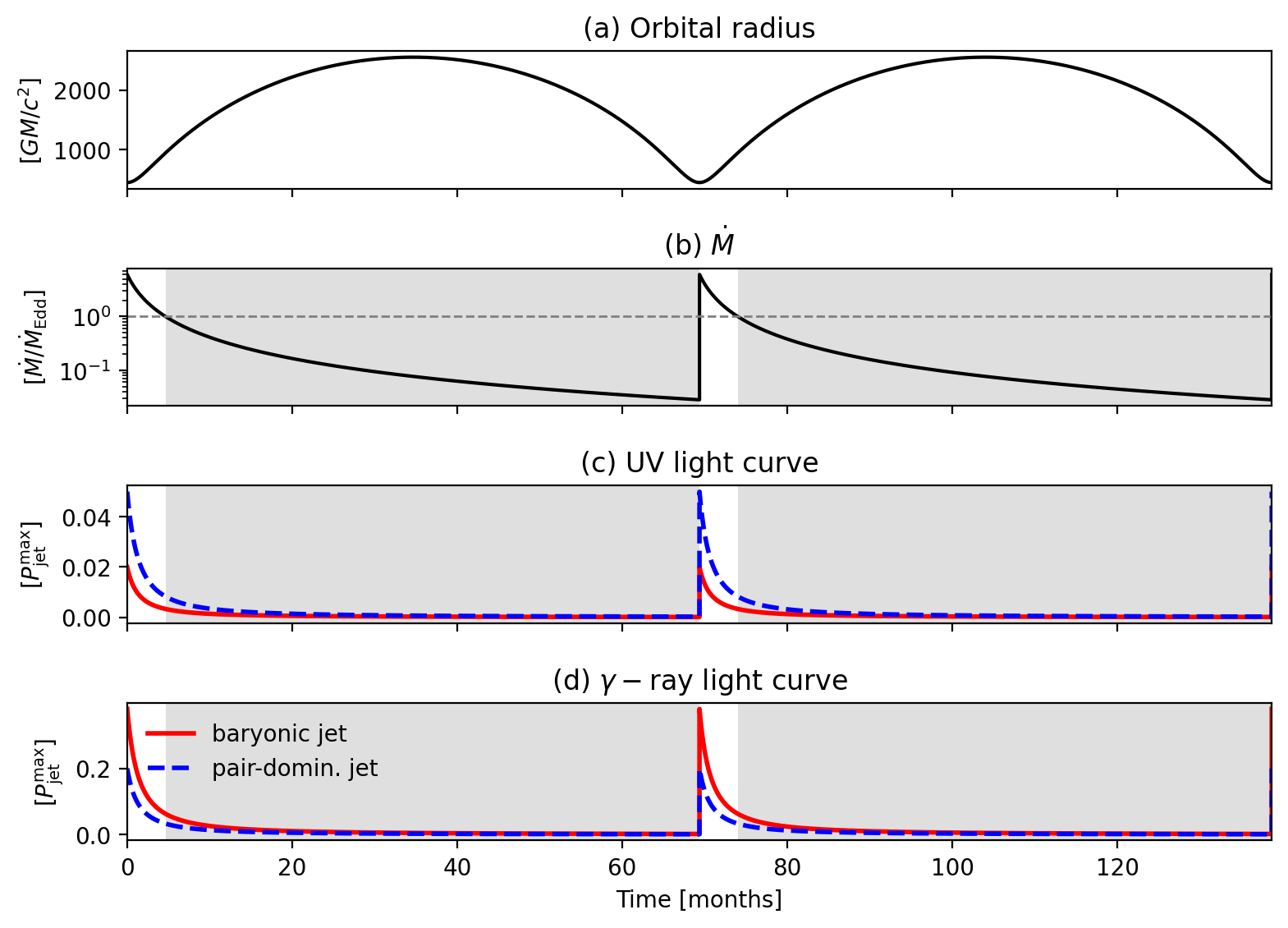}
\caption{
{Temporal} 
 {evolution} 
 of orbital-dependent emission properties over two orbital periods of a demonstrative  EMRB  with the parameters indicated by magenta dots in Figures \ref{fig:period}, \ref{fig:pericenter}, and \ref{fig:Mcap_500} (see also Table \ref{tab:parameters} for the list of parameters). The miniquasar reaches pericenter at $t = 0$, triggering a transient accretion episode and the onset of jet activity. 
(\textbf{a}) Orbital radius as a function of time, Equation~(\ref{eq:r(t)}). 
(\textbf{b})  Eddington accretion rate of the miniquasar in logarithmic scale. The horizontal dashed line indicates the value $\dot{M}_{\rm mq}/\dot{M}_{\rm Edd}=1$. As the jet is assumed to be launched when the accretion is super-Eddington, Equation (\ref{eq:mdot_constraint1}), the shaded gray regions indicate where the jet is expected to be quenched.
(\textbf{c}) Light curve of UV power  from the interaction between the miniquasar jet and SMBH accretion disk, in units of  $P_{\rm jet}^{\rm max}$,  where $P_{\rm jet}^{\rm max}$ is the jet power at $t=0$.
(\textbf{d}) Light curve of  gamma-ray power  from the jet--disk interaction, in units of  $P_{\rm jet}^{\rm max}$.    In panels (\textbf{c},\textbf{d}), the emission power are expected to diminish to zero as a result of jet-quenching.
The case for a pair-dominated jet $(\xi_{\rm e^{+}}, \xi_{\rm p}) = (1, 0)$ and a baryonic jet $(\xi_{\rm e^{+}}, \xi_{\rm p}) = (0, 1)$ is shown by, respectively, the solid red and dashed blue profiles. The adopted parameters are summarized in \mbox{Table \ref{tab:parameters}.}
}\label{fig:lc}
\end{figure}

By adopting the emission-related parameters summarized in Table~\ref{tab:parameters}, we examine two representative jet compositions: a pair-dominated jet ($\xi_{{\rm e}^{+}} = 1$, $\xi_{{\rm p}} = 0$) and a baryonic jet ($\xi_{{\rm e}^{+}} = 0$, $\xi_{{\rm p}} = 1$) in { panels (c) and (d) in} Figure \ref{fig:lc}. It is assumed that 50$\%$ of the lepton energy is converted into emissions, $\mathcal{Q}=0.5$. Additionally, we adopt $f_{\rm IC}=0.8\mathcal{Q}$ (so ${U_{\rm ph}}/{U_{\rm B}}=4$ to align with the expectation that there will be substantial IC emissions due to the effective amplification of the energy of the external photon from the SMBH accretion flow by the relativistic photons present within the jet.  The corresponding model light curves are calculated as described in Sections~\ref{sec:modeling_emisssion} and~\ref{sec:two-stage}.

Although the results quantitatively depend on the chosen values of the parameters, qualitative trends reveal meaningful insights into how the emissions from pair-dominated and baryonic jets are modulated by orbital dynamics. 
The UV and gamma-ray power  from the jet--disk interaction, displayed in panels (c) and (d) of Figure \ref{fig:lc}, respectively, are shown in units of the maximum jet power  corresponding to the initial accretion rate at $t=0$. In addition, the jet power evolution is proportional to the accretion, as described in \mbox{Equation (\ref{eq:mdot_t}).} These profiles are obtained through observing the way the intrisic jet power is carried by different particle species (Section \ref{sec: jet_composition}), how the carried power could be converted to  radiation power, and the associated conversion efficiency at different frequencies\mbox{ (Sections \ref{sec:lep_radiation} and \ref{sec:had_radiation}).}  
For the pair-dominated jet, the gamma rays are contributed by the IC process via leptons. In comparison, the gamma rays are contributed by neutral pion decay and the IC process for both primary and secondary leptons. 
For both types of jet, the UV emissions comprise the synchrotron emissions from leptons. 
Note that the gray regions shown in the panels of Figure \ref{fig:lc} indicate the system is sub-Eddinton and the jet is quenched. Consequently, in panels (c) and (d) of Figure \ref{fig:lc}, it is expected that the calculated emission profiles located within the gray-shaded area are expected to diminish to zero as a result of jet quenching. This observed discontinuity in the multi-frequency light curve profiles of electromagnetic radiation from EMRBs might therefore offer empirical evidence for the activity of miniquasar jets.

The pair-dominated jet has a relatively stronger UV power, and the baryonic jet instead has a relatively larger gamma-ray power. For the adopted ratio $\langle\gamma_{\rm p}\rangle/\langle\gamma_{\rm e}\rangle=10^{-2}$, the power  ratios are $P_{\gamma{\rm-ray}}/P_{\rm UV}=4$  and $P_{\gamma{\rm-ray}}/P_{\rm UV}\sim20$ for the pair-dominated jet and the baryonic jet, as indicated in Figure \ref{fig:ratios_compare} and discussed in Section \ref{sec:ratio_estimation}. 
The power of multi-wavelength emissions is transformed from the power contained within the leptonic and hadronic components of the jet, and a reduction in $\langle\gamma_{\rm p}\rangle/\langle\gamma_{\rm e}\rangle$ indicates that more power is carried by protons, resulting in increased gamma-ray emissions by efficient pp collision.  The presence of a hadronic component significantly enhances the potential of the  EMRB system to appear as a bright transient gamma-ray source.  


\section{Summary and Final Remarks}\label{S4}
Motivated by the recognition that quasi-steady-state conditions in black hole jets may obscure the diagnostic temporal variability that is essential for constraining jet composition, in this work, we investigate the emission properties of  EMRBs, wherein a miniquasar orbits an SMBH. In the configurations considered, the pericenter and apocenter of the miniquasar orbit reside within and beyond the outer boundary of the geometrically thin accretion disk of SMBH, respectively (see also Section \ref{sec:orbital}). This orbital structure induces a sequence of distinctive dynamical phenomena, including periodic penetration of the SMBH disk, episodic accretion onto the miniquasar, and subsequent jet-launching events. Furthermore, the jet emitted by the miniquasar may interact directly with the accretion environment of the SMBH, producing collision-driven high-energy signatures.

Compared to the relativistic jets in AGN, any baryonic component within the miniquasar jet is expected to achieve significantly lower maximum energies. As a consequence, hadronic interactions arising from jet--disk collisions are dominated by proton--proton collisions rather than proton--photon processes. 
It is demonstrated (Section \ref{sec:lc}) that,  during the jet--disk interaction, the resulting multi-wavelength light curves encode characteristic temporal and spectral imprints of both leptonic and hadronic emission components. By integrating variability timescales, jet power evolution, and multi-wavelength radiation features, it is suggested that these systems offer a novel laboratory for constraining the composition of black hole jets.

The key results of the work are summarized as follows:
\begin{itemize}

\item The period of the EMRBs of interest here is on the order of years (see also Figure \ref{fig:period}). Recurrent encounters between the miniquasar jet and the SMBH accretion flow lead to periodic accretion episodes and jet activity from the miniquasar, characterized by a two-stage accretion process described in Section \ref{sec:two-stage}.

\item In the first stage, when the miniquasar traverses the thin accretion disk of the SMBH, it captures material through Bondi accretion. 
In the second stage, the captured material is then gradually funneled into the miniquasar,  usually leading to super-Eddington accretion. The jet is assumed to be launching when the super-Eddinton accretion onto the miniquasar is satisfied, Equation (\ref{eq:mdot_constraint1}). Typical intrinsic jet power is about $10^{42-44}~{\rm erg/s^{-1}}$ (see the estimation described after Equation (\ref{eq:pjet})).

The duration of both the accretion phase and the associated jet activity is determined primarily by the viscous timescale of the inflowing material. As the captured material is gradually depleted, both the accretion rate and the resulting jet power decrease over time  before the jet activity is quenched.

\item Once the miniquasar jet is launched and interacts with the SMBH’s accretion disk, the resulting emission signatures serve as a valuable diagnostic for probing the jet composition. The multi-frequency radiation power from the transient jet can be estimated based on the jet power carried by different particle species and their respective energy-conversion efficiencies to radiation (see also Sections \ref{sec:lep_rt},  \ref{sec:had_rt},  \ref{sec:lep_radiation} and  \ref{sec:had_radiation}). 

\item The leptonic components within the jet are the source of optical/UV emissions through synchrotron radiation and approximate MeV/GeV gamma rays via IC scattering. In contrast, the proton--proton interactions from the hadronic constituents can yield MeV/GeV gamma-rays through either neutral pion decay or the IC process involving secondary leptons generated by charged pion decay. Additionally, synchrotron radiation from these secondary leptons may also contribute to optical/UV emissions. An analysis of the ratios of different frequency emissions is provided in \mbox{Section \ref{sec:ratio_estimation}.}

\item Among the diverse multi-frequency features associated with the  EMRBs under consideration, we focus primarily on emissions from  miniquasar jets and their interaction with the SMBH accretion flow. In Section \ref{sec:lc}, we model the orbital dynamics and corresponding light curves of UV and gamma-ray power resulting from jet--disk collisions for a demonstration case. With the efficient pp interaction, \mbox{Equation (\ref{eq:fpp=1}),} and the radiation conversion rate, \mbox{Equation (\ref{eq:pion0decay}),} during the jet--disk interaction, the ratio of gamma-ray-to-UV emission serves as a valuable diagnostic to identify the presence of hadronic components within the jet. This ratio increases with the fraction of jet power carried by hadrons. 
\end{itemize}

Other emission components, such as radiation from the SMBH's thin accretion disk, may further shape the observed spectral energy distribution, particularly in the optical and ultraviolet bands. However, the time variation of the UV light curve and the profile of the spectra may help distengle the sources of UV emissions.
Moreover, recurrent interactions between the miniquasar jet and the SMBH accretion flow may induce
variations in disk emissions, providing an additional avenue for identifying  EMRBs. The transient accretion flow around the miniquasar itself
could also produce detectable X-ray emission. 

 In this work, we estimate the emission properties without considering the full energy distribution of each particle species. We also explore how the multi-frequency power evolves over time as the accretion rate declines, neglecting for now the influence of relative motion between the system and the observer. More detailed spectral modeling that accounts for these effects will be presented in future work.
 In addition, to further advance our understanding of the complex physical processes involved, it is essential to pursue comprehensive {modeling} 
 that properly accounts for the transient accretion episodes and the corresponding jet-launching mechanisms.

Complementing the electromagnetic signatures, neutrino production via pion decay will offer a useful means to verify the presence of hadronic interaction, and hence provide information regarding the hadronic content in relativistic jets of accreting black holes.  A rough estimate of the resulting neutrino power can be obtained using (cf. Equation (\ref{eq:hadron_to_gamma_GeV}))
\begin{equation}\label{eq:hadron_to_nurtino}
    P_{{\rm p}\to \nu}\approx f_{\rm pp}(\dfrac{2}{3}f_{\pi^{\pm}\to \nu})P_{\rm p}\;.
\end{equation}

{With} $f_{\pi^{\pm}\to \nu}\sim 10\%$ and a modest intrinsic jet power $P_{\rm jet}\sim 10^{42-44}~{\rm erg/s^{-1}}$, the intrinsic neutrino power due to the interaction between the miniquasar baryonic jet and the SMBH disk can produce $P_{{\rm p}\to \nu}\sim 10^{40-42}~\rm erg/s$. For an EMRB system at a distance of \mbox{$\sim$$100$ Mpc,} the corresponding neutrino flux would be $\sim$$10^{-14}$--$10^{-12}~{\rm erg/s/cm^{2}}$, potentially accessible to neutrino detectors such as IceCube \citep[][]{icecube2015}. We note that neutrino fluxes associated with the miniquasar jet are expected to have lower energy\linebreak   ($\sim$GeV-TeV) compared to neutrinos produced via the SMBH jet, e.g., TXS 0506+056 \cite[][]{icecube2018,abb2024}.

The detection of high-energy ($\geq$10 GeV–PeV) astrophysical neutrinos is primarily pursued by large-volume observatories such as IceCube \citep[][]{icecube2017}, KM3NeT/ARCA \cite[][]{KM3NET2021,KM3NET2025}, Baikal-GVD \citep[][]{baikal2018}, and future facilities including IceCube-Gen2 \citep[][]{icecube-gen2} and P-ONE \citep[][]{pone2019,pone2020}, which are designed to capture hadronic processes in energetic transients.
Meanwhile, GeV-sensitive detectors such as Deep Underground Neutrino Experiment (DUNE) \citep[][]{dune2020}, the future  Hyper-Komiokande (Hyper-K) \citep[][]{hyperK2018}, and the Oscillation Research with Cosmics in the Abyss (ORCA) in KM3NeT \citep[][]{ORCA2024} can offer valuable complementary energy coverage that may help constrain low-energy extensions of these neutrino spectra in exceptional nearby or high-fluence events.
Together with next-generation electromagnetic facilities, including
the Thirty Meter Telescope (TMT) \citep{TMT2013} and Extremely Large Telescope (ELT) \citep{EELT2018,ELT2023}
in the optical and infrared wavebands, 
Advanced Telescope for High Energy Astrophysics (ATHENA) \citep{Athena2012,Athena2020}
and Imaging and Spectroscopy Mission (XRISM) 
\citep[][]{xrism2020}
in the X-ray bands, and
the  gamma-ray telescopes succeeding
Fermi Gamma-ray Space Telescope \citep[][]{Fermi2022}, 
and the facilities capable of detecting neutrinos at GeV energies,  these observatories will form an increasingly synergistic multi-messenger network for probing jet composition and orbital-phase-dependent emissions in EMRB systems.




\vspace{6pt}

\funding{This work is supported by the Yushan Fellow Program of the Ministry of Education (MoE) of Taiwan (ROC), the National Science and Technology Council (NSTC) of Taiwan (ROC) under the grant 112-2112-M-003-010-MY3.}

\dataavailability{Data is contained within the article.} 

\acknowledgments{The author gratefully acknowledges Kinwah Wu and Ellis Owen for their stimulating discussions, and the anonymous referees for their helpful comments that significantly improved the article.
The author also thanks Kaye Jiale Li and Afif Ismail for careful proofreading and Hayden Ping Hei Ng for providing helpful references. This work has used the NASA Astrophysics Data System.}

\conflictsofinterest{The authors declare no conflicts of interest.} 





\appendixtitles{yes} 
\appendixstart
\appendix

\section{Mass Accretion and Accretion Rate of the Miniquasar}\label{app:Mcap_mq}
Here, we qualitatively estimate the accreted mass and the initial accretion rate, according to Equations (\ref{eq:m_cap}) and (\ref{eq:mdot_norm}). 

Assuming that the miniquasar passes the SMBH disk at approximately the pericenter $r_{\rm peri}$, it is assumed that
\begin{equation}
    v_{\rm rel}\sim v_{\rm orb}=(\frac{{\rm G}M_{\rm SMBH}}{r_{\rm peri}})^{1/2}\;,
\end{equation}
and 
\begin{equation}
    v_{\rm rel} \times t_{\rm pass}= v_{\rm orb}\times \frac{H}{v_{\rm orb}}=H\sim 0.01~r_{\rm peri}\;.
\end{equation}

\begin{figure}[H]
\includegraphics[width=0.8\textwidth]{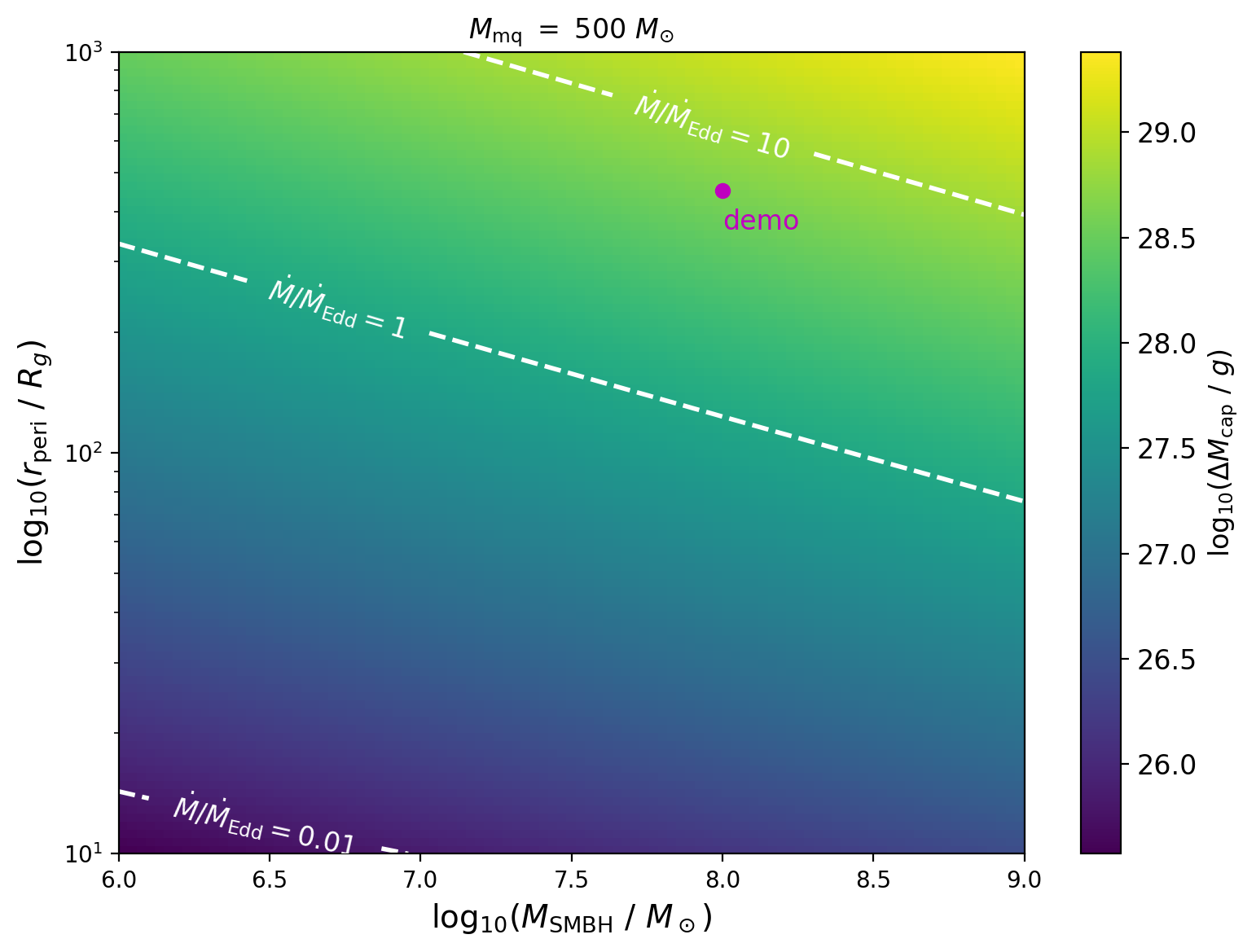}
\caption{{Mass} 
 captured by a 
$500~M_{\odot}$ miniquasar during its passage through the accretion disk of an SMBH of mass 
$M_{\rm SMBH}$, evaluated at the   $r_{\rm peri}$.
The selected parameters for our demonstrative case is indicated by the magenta dot. Contours of constant accretion rate are overlapped by the dashed lines.
\label{fig:Mcap_500}}
\end{figure}

{In} addition, we adopted the semi-analytical solution for the disk density \citep[][]{Shakura1973,kato1998}
\begin{equation}
     \rho=8~\alpha^{-7/10} (\dfrac{M_{\rm SMBH}}{M_{\odot}})^{-7/10} (\frac{\dot{M}}{\dot{M}_{\rm Edd}})^{2/5} (\frac{r_{\rm peri}}{2R_{\rm g}})^{-33/20} f^{2/5}\;,
\end{equation}
where $f\equiv 1-(6R_{\rm g}/r_{\rm peri})^{1/2}$ and the typical values for viscosity $\alpha=0.01$ and accretion rate $\dot{M}=0.5\dot{M}_{\rm Edd}$ are adopted. 
The computed captured mass for $M_{\rm mq}=500M_{\odot}$ with different $M_{\rm SMBH}$ and $r_{\rm peri}$ is shown in Figure \ref{fig:Mcap_500}. 

The accretion rate of the miniquasar is related to $t_{\rm visc}$. Assuming $t_{\rm visc}=2$ months, the initial accretion rate is calculated using Equation (\ref{eq:mdot_norm}) and is indicated by the dashed contours in the figure.

\begin{adjustwidth}{-\extralength}{0cm}
\printendnotes[custom]

\reftitle{References}
\PublishersNote{}
\end{adjustwidth}

\begin{thebibliography}{999}

\bibitem[{Fender} et~al.(2004){Fender}, {Belloni}, and {Gallo}]{Fender2004}
{Fender}, R.P.; {Belloni}, T.M.; {Gallo}, E.
\newblock {Towards a unified model for black hole X-ray binary jets}.
\newblock {\em Mon. Not. R. Astron. Soc.} {\bf 2004}, {\em 355}.~1105--1118,
\newblock {\url{https://doi.org/10.1111/j.1365-2966.2004.08384.x}}.

\bibitem[{Corbel}(2011)]{cor2011}
{Corbel}, S.
\newblock {Microquasars: An observational review.}
\newblock In \textit{Proceedings of the Jets at All Scales, IAU Symposium;} 
 {Romero}, G.E., {Sunyaev},
  R.A., {Belloni}, T., Eds.;  {Cambridge University Press: Cambridge, UK,} 
 2011; Volume 275, pp.
  205--214.
\newblock {\url{https://doi.org/10.1017/S1743921310016054}}.

\bibitem[{Kylafis} et~al.(2012){Kylafis}, {Contopoulos}, {Kazanas}, and
  {Christodoulou}]{kylasfis2012}
{Kylafis}, N.D.; {Contopoulos}, I.; {Kazanas}, D.; {Christodoulou}, D.M.
\newblock {Formation and destruction of jets in X-ray binaries}.
\newblock {\em Astron. Astrophys.} {\bf 2012}, {\em 538},~A5.
\newblock {\url{https://doi.org/10.1051/0004-6361/201117052}}.

\bibitem[{Trump} et~al.(2011){Trump}, {Impey}, {Kelly}, {Civano}, {Gabor},
  {Diamond-Stanic}, {Merloni}, {Urry}, {Hao}, {Jahnke}, {Nagao}, {Taniguchi},
  {Koekemoer}, {Lanzuisi}, {Liu}, {Mainieri}, {Salvato}, and
  {Scoville}]{trump2011}
{Trump}, J.R.; {Impey}, C.D.; {Kelly}, B.C.; {Civano}, F.; {Gabor}, J.M.;
  {Diamond-Stanic}, A.M.; {Merloni}, A.; {Urry}, C.M.; {Hao}, H.; {Jahnke}, K.;
   et~al.
\newblock {Accretion Rate and the Physical Nature of Unobscured Active
  Galaxies}.
\newblock {\em Astrophys. J.} {\bf 2011}, {\em 733},~60.
\newblock {\url{https://doi.org/10.1088/0004-637X/733/1/60}}.

\bibitem[{Fabian}(2012)]{fabian2012}
{Fabian}, A.C.
\newblock {Observational Evidence of Active Galactic Nuclei Feedback}.
\newblock {\em Annu. Rev. Astron. Astrophys. } {\bf 2012}, {\em 50},~455--489.
\newblock {\url{https://doi.org/10.1146/annurev-astro-081811-125521}}.

\bibitem[{Hovatta} and {Lindfors}(2019)]{Hov2019}
{Hovatta}, T.; {Lindfors}, E.
\newblock {Relativistic Jets of Blazars}.
\newblock {\em New Astron. Rev.} {\bf 2019}, {\em 87},~101541.
\newblock {\url{https://doi.org/10.1016/j.newar.2020.101541}}.

\bibitem[{Combes}(2021)]{combes2021}
{Combes}, F.
\newblock {\em {Active Galactic Nuclei: Fueling and Feedback}};  {IoP Publishing:  Bristol, UK,} 
 2021.
\newblock {\url{https://doi.org/10.1088/2514-3433/ac2a27}}.

\bibitem[{Asada} and {Nakamura}(2012)]{asada2012}
{Asada}, K.; {Nakamura}, M.
\newblock {The Structure of the M87 Jet: A Transition from Parabolic to Conical
  Streamlines}.
\newblock {\em Astrophys. J. Lett.} {\bf 2012}, {\em 745},~L28.
\newblock {\url{https://doi.org/10.1088/2041-8205/745/2/L28}}.

\bibitem[{Kravchenko} et~al.(2025){Kravchenko}, {Pashchenko}, {Homan},
  {Kovalev}, {Lister}, {Pushkarev}, {Ros}, and {Savolainen}]{kra2025}
{Kravchenko}, E.V.; {Pashchenko}, I.N.; {Homan}, D.C.; {Kovalev}, Y.Y.;
  {Lister}, M.L.; {Pushkarev}, A.B.; {Ros}, E.; {Savolainen}, T.
\newblock {MOJAVE---XXII. Brightness temperature distributions and geometric
  profiles along parsec-scale active galactic nucleus jets}.
\newblock {\em Mon. Not. R. Astron. Soc.} {\bf 2025}, {\em 538},~2008--2030.
\newblock {\url{https://doi.org/10.1093/mnras/staf343}}.

\bibitem[{Asada} et~al.(2014){Asada}, {Nakamura}, {Doi}, {Nagai}, and
  {Inoue}]{asada2014}
{Asada}, K.; {Nakamura}, M.; {Doi}, A.; {Nagai}, H.; {Inoue}, M.
\newblock {Discovery of Sub- to Superluminal Motions in the M87 Jet: An
  Implication of Acceleration from Sub-relativistic to Relativistic Speeds}.
\newblock {\em Astrophys. J. Lett.} {\bf 2014}, {\em 781},~L2.
\newblock {\url{https://doi.org/10.1088/2041-8205/781/1/L2}}.

\bibitem[{Marin} et~al.(2024){Marin}, {Barnouin}, {Wu}, and
  {Lopez-Rodriguez}]{marin2024}
{Marin}, F.; {Barnouin}, T.; {Wu}, K.; {Lopez-Rodriguez}, E.
\newblock {Forgotten treasures in the HST/FOC UV imaging polarimetric archives
  of active galactic nuclei: III. Five years monitoring of M87}.
\newblock {\em Astron. Astrophys. } {\bf 2024}, {\em 692},~A179.
\newblock {\url{https://doi.org/10.1051/0004-6361/202451390}}.

\bibitem[{Fan} et~al.(2018){Fan}, {Wu}, and {Liao}]{fan2018}
{Fan}, X.L.; {Wu}, Q.; {Liao}, N.H.
\newblock {Constraints on the Composition, Magnetization, and Radiative
  Efficiency in the Jets of Blazars}.
\newblock {\em Astrophys. J. } {\bf 2018}, {\em 861},~97.
\newblock {\url{https://doi.org/10.3847/1538-4357/aac959}}.

\bibitem[{Owen} and {Yang}(2022{\natexlab{a}})]{owen2022_lep}
{Owen}, E.R.; {Yang}, H.Y.K.
\newblock {Multiwavelength emission from leptonic processes in ageing galaxy
  bubbles}.
\newblock {\em Mon. Not. R. Astron. Soc.} {\bf 2022}, {\em 510},~5834--5853.
\newblock {\url{https://doi.org/10.1093/mnras/stac119}}.

\bibitem[{Owen} and {Yang}(2022{\natexlab{b}})]{owen2022}
{Owen}, E.R.; {Yang}, H.Y.K.
\newblock {Emission from hadronic and leptonic processes in galactic jet-driven
  bubbles}.
\newblock {\em Mon. Not. R. Astron. Soc.} {\bf 2022}, {\em 516},~1539--1556.
\newblock {\url{https://doi.org/10.1093/mnras/stac2289}}.

\bibitem[{Cerruti}(2020)]{Matteo2020}
{Cerruti}, M.
\newblock {Leptonic and Hadronic Radiative Processes in Supermassive-Black-Hole
  Jets}.
\newblock {\em Galaxies} {\bf 2020}, {\em 8},~72.
\newblock {\url{https://doi.org/10.3390/galaxies8040072}}.

\bibitem[{Lin} et~al.(2023){Lin}, {Yang}, and {Owen}]{lin2023}
{Lin}, Y.H.; {Yang}, H.Y.K.; {Owen}, E.R.
\newblock {Evolution and feedback of AGN jets of different cosmic ray
  composition}.
\newblock {\em Mon. Not. R. Astron. Soc.} {\bf 2023}, {\em 520},~963--975.
\newblock {\url{https://doi.org/10.1093/mnras/stad185}}.

\bibitem[{Kantzas} et~al.(2023){Kantzas}, {Markoff}, {Lucchini}, {Ceccobello},
  and {Chatterjee}]{kan2023}
{Kantzas}, D.; {Markoff}, S.; {Lucchini}, M.; {Ceccobello}, C.; {Chatterjee},
  K.
\newblock {Exploring the role of composition and mass loading on the properties
  of hadronic jets}.
\newblock {\em Mon. Not. R. Astron. Soc.} {\bf 2023}, {\em 520},~6017--6039.
\newblock {\url{https://doi.org/10.1093/mnras/stad521}}.

\bibitem[{Rodrigues} et~al.(2024){Rodrigues}, {Paliya}, {Garrappa}, {Omeliukh},
  {Franckowiak}, and {Winter}]{Rodrigues2024}
{Rodrigues}, X.; {Paliya}, V.S.; {Garrappa}, S.; {Omeliukh}, A.; {Franckowiak},
  A.; {Winter}, W.
\newblock {Leptohadronic multi-messenger modeling of 324 gamma-ray blazars}.
\newblock {\em Astron. Astrophys. } {\bf 2024}, {\em 681},~A119.
\newblock {\url{https://doi.org/10.1051/0004-6361/202347540}}.

\bibitem[{Lico} et~al.(2022){Lico}, {Casadio}, {Jorstad}, {G{\'o}mez},
  {Marscher}, {Traianou}, {Kim}, {Zhao}, {Fuentes}, {Cho}, {Krichbaum},
  {Hervet}, {O'Brien}, {Boccardi}, {Myserlis}, {Agudo}, {Alberdi}, {Weaver},
  and {Zensus}]{lico2022}
{Lico}, R.; {Casadio}, C.; {Jorstad}, S.G.; {G{\'o}mez}, J.L.; {Marscher},
  A.P.; {Traianou}, E.; {Kim}, J.Y.; {Zhao}, G.Y.; {Fuentes}, A.; {Cho}, I.;
  et~al.
\newblock {New jet feature in the parsec-scale jet of the blazar OJ 287
  connected to the 2017 teraelectronvolt flaring activity}.
\newblock {\em Astron. Astrophys. } {\bf 2022}, {\em 658},~L10.
\newblock {\url{https://doi.org/10.1051/0004-6361/202142948}}.

\bibitem[{Valtonen} et~al.(2024){Valtonen}, {Zola}, {Gupta}, {Kishore},
  {Gopakumar}, {Jorstad}, {Wiita}, {Gu}, {Nilsson}, {Marscher}, {Zhang},
  {Hudec}, {Matsumoto}, {Drozdz}, {Ogloza}, {Berdyugin}, {Reichart},
  {Mugrauer}, {Dey}, {Pursimo}, {Lehto}, {Ciprini}, {Nakaoka}, {Uemura},
  {Imazawa}, {Zejmo}, {Kouprianov}, {Davidson}, {Sadun}, {{\v{S}}trobl},
  {Weaver}, and {Jel{\'\i}nek}]{val2024}
{Valtonen}, M.J.; {Zola}, S.; {Gupta}, A.C.; {Kishore}, S.; {Gopakumar}, A.;
  {Jorstad}, S.G.; {Wiita}, P.J.; {Gu}, M.; {Nilsson}, K.; {Marscher}, A.P.;
  et~al.
\newblock {Evidence of Jet Activity from the Secondary Black Hole in the OJ 287
  Binary System}.
\newblock {\em Astrophys. J. Lett.} {\bf 2024}, {\em 968},~L17.
\newblock {\url{https://doi.org/10.3847/2041-8213/ad4d9b}}.

\bibitem[{Zhang} et~al.(1997){Zhang}, {Ebisawa}, {Sunyaev}, {Ueda}, {Harmon},
  {Sazonov}, {Fishman}, {Inoue}, {Paciesas}, and {Takahash}]{zheng1997}
{Zhang}, S.N.; {Ebisawa}, K.; {Sunyaev}, R.; {Ueda}, Y.; {Harmon}, B.A.;
  {Sazonov}, S.; {Fishman}, G.J.; {Inoue}, H.; {Paciesas}, W.S.; {Takahash}, T.
\newblock {Broadband High-Energy Observations of the Superluminal Jet Source
  GRO J1655-40 during an Outburst}.
\newblock {\em Astrophys. J. } {\bf 1997}, {\em 479},~381--387.
\newblock {\url{https://doi.org/10.1086/303870}}.

\bibitem[{Kaaret} et~al.(2003){Kaaret}, {Corbel}, {Tomsick}, {Fender},
  {Miller}, {Orosz}, {Tzioumis}, and {Wijnands}]{kaa2003}
{Kaaret}, P.; {Corbel}, S.; {Tomsick}, J.A.; {Fender}, R.; {Miller}, J.M.;
  {Orosz}, J.A.; {Tzioumis}, A.K.; {Wijnands}, R.
\newblock {X-Ray Emission from the Jets of XTE J1550-564}.
\newblock {\em Astrophys. J. } {\bf 2003}, {\em 582},~945--953.
\newblock {\url{https://doi.org/10.1086/344540}}.

\bibitem[{Tagawa} et~al.(2023{\natexlab{a}}){Tagawa}, {Kimura}, {Haiman},
  {Perna}, and {Bartos}]{tagawa2023a}
{Tagawa}, H.; {Kimura}, S.S.; {Haiman}, Z.; {Perna}, R.; {Bartos}, I.
\newblock {Observable Signature of Merging Stellar-mass Black Holes in Active
  Galactic Nuclei}.
\newblock {\em Astrophys. J. } {\bf 2023}, {\em 950},~13.
\newblock {\url{https://doi.org/10.3847/1538-4357/acc4bb}}.

\bibitem[{Tagawa} et~al.(2023{\natexlab{b}}){Tagawa}, {Kimura}, and
  {Haiman}]{tagawa2023}
{Tagawa}, H.; {Kimura}, S.S.; {Haiman}, Z.
\newblock {High-energy Electromagnetic, Neutrino, and Cosmic-Ray Emission by
  Stellar-mass Black Holes in Disks of Active Galactic Nuclei}.
\newblock {\em Astrophys. J. } {\bf 2023}, {\em 955},~23.
\newblock {\url{https://doi.org/10.3847/1538-4357/ace71d}}.

\bibitem[{Kocsis} et~al.(2011){Kocsis}, {Yunes}, and {Loeb}]{kocsis2011}
{Kocsis}, B.; {Yunes}, N.; {Loeb}, A.
\newblock {Observable signatures of extreme mass-ratio inspiral black hole
  binaries embedded in thin accretion disks}.
\newblock {\em Phys. Rev. D Part. Fields Gravit. Cosmol. } {\bf 2011}, {\em 84},~024032.
\newblock {\url{https://doi.org/10.1103/PhysRevD.84.024032}}.

\bibitem[{Lehto} and {Valtonen}(1996)]{Lehto1996}
{Lehto}, H.J.; {Valtonen}, M.J.
\newblock {OJ 287 Outburst Structure and a Binary Black Hole Model}.
\newblock {\em Astrophys. J. } {\bf 1996}, {\em 460},~207.
\newblock {\url{https://doi.org/10.1086/176962}}.

\bibitem[{Pihajoki}(2016)]{pihajoki2016}
{Pihajoki}, P.
\newblock {Black hole accretion disc impacts}.
\newblock {\em Mon. Not. R. Astron. Soc.} {\bf 2016}, {\em 457},~1145--1161.
\newblock {\url{https://doi.org/10.1093/mnras/stv3023}}.

\bibitem[{McLaughlin}(2013)]{mclaughlin2013}
{McLaughlin}, M.A.
\newblock {The North American Nanohertz Observatory for Gravitational Waves}.
\newblock {\em Class. Quantum Gravity} {\bf 2013}, {\em 30},~224008.
\newblock {\url{https://doi.org/10.1088/0264-9381/30/22/224008}}.

\bibitem[{Kramer} and {Champion}(2013)]{kramer2013}
{Kramer}, M.; {Champion}, D.J.
\newblock {The European Pulsar Timing Array and the Large European Array for
  Pulsars}.
\newblock {\em Class. Quantum Gravity} {\bf 2013}, {\em 30},~224009.
\newblock {\url{https://doi.org/10.1088/0264-9381/30/22/224009}}.

\bibitem[{Manchester} et~al.(2013){Manchester}, {Hobbs}, {Bailes}, {Coles},
  {van Straten}, {Keith}, {Shannon}, {Bhat}, {Brown}, {Burke-Spolaor},
  {Champion}, {Chaudhary}, {Edwards}, {Hampson}, {Hotan}, {Jameson}, {Jenet},
  {Kesteven}, {Khoo}, {Kocz}, {Maciesiak}, {Oslowski}, {Ravi}, {Reynolds},
  {Sarkissian}, {Verbiest}, {Wen}, {Wilson}, {Yardley}, {Yan}, and
  {You}]{manchester2013}
{Manchester}, R.N.; {Hobbs}, G.; {Bailes}, M.; {Coles}, W.A.; {van Straten},
  W.; {Keith}, M.J.; {Shannon}, R.M.; {Bhat}, N.D.R.; {Brown}, A.;
  {Burke-Spolaor}, S.G.;  et~al.
\newblock {The Parkes Pulsar Timing Array Project}.
\newblock {\em Publ. Astron. Soc. Aust. } {\bf 2013}, {\em 30},~e017.
\newblock {\url{https://doi.org/10.1017/pasa.2012.017}}.

\bibitem[{Verbiest} et~al.(2016){Verbiest}, {Lentati}, {Hobbs}, {van
  Haasteren}, {Demorest}, {Janssen}, {Wang}, {Desvignes}, {Caballero}, {Keith},
  {Champion}, {Arzoumanian}, {Babak}, {Bassa}, {Bhat}, {Brazier}, {Brem},
  {Burgay}, {Burke-Spolaor}, {Chamberlin}, {Chatterjee}, {Christy}, {Cognard},
  {Cordes}, {Dai}, {Dolch}, {Ellis}, {Ferdman}, {Fonseca}, {Gair},
  {Garver-Daniels}, {Gentile}, {Gonzalez}, {Graikou}, {Guillemot}, {Hessels},
  {Jones}, {Karuppusamy}, {Kerr}, {Kramer}, {Lam}, {Lasky}, {Lassus},
  {Lazarus}, {Lazio}, {Lee}, {Levin}, {Liu}, {Lynch}, {Lyne}, {Mckee},
  {McLaughlin}, {McWilliams}, {Madison}, {Manchester}, {Mingarelli}, {Nice},
  {Os{\l}owski}, {Palliyaguru}, {Pennucci}, {Perera}, {Perrodin}, {Possenti},
  {Petiteau}, {Ransom}, {Reardon}, {Rosado}, {Sanidas}, {Sesana}, {Shaifullah},
  {Shannon}, {Siemens}, {Simon}, {Smits}, {Spiewak}, {Stairs}, {Stappers},
  {Stinebring}, {Stovall}, {Swiggum}, {Taylor}, {Theureau}, {Tiburzi},
  {Toomey}, {Vallisneri}, {van Straten}, {Vecchio}, {Wang}, {Wen}, {You},
  {Zhu}, and {Zhu}]{verbiest2016}
{Verbiest}, J.P.W.; {Lentati}, L.; {Hobbs}, G.; {van Haasteren}, R.;
  {Demorest}, P.B.; {Janssen}, G.H.; {Wang}, J.B.; {Desvignes}, G.;
  {Caballero}, R.N.; {Keith}, M.J.;  et~al.
\newblock {The International Pulsar Timing Array: First data release}.
\newblock {\em Mon. Not. R. Astron. Soc.} {\bf 2016}, {\em 458},~1267--1288.
\newblock {\url{https://doi.org/10.1093/mnras/stw347}}.

\bibitem[{Pan} and {Yang}(2021)]{pna2021}
{Pan}, Z.; {Yang}, H.
\newblock {Formation rate of extreme mass ratio inspirals in active galactic
  nuclei}.
\newblock {\em Phys. Rev. D Part. Fields Gravit. Cosmol. } {\bf 2021}, {\em 103},~103018.
\newblock {\url{https://doi.org/10.1103/PhysRevD.103.103018}}.

\bibitem[{Klein} et~al.(2016){Klein}, {Barausse}, {Sesana}, {Petiteau},
  {Berti}, {Babak}, {Gair}, {Aoudia}, {Hinder}, {Ohme}, and
  {Wardell}]{klein2016}
{Klein}, A.; {Barausse}, E.; {Sesana}, A.; {Petiteau}, A.; {Berti}, E.;
  {Babak}, S.; {Gair}, J.; {Aoudia}, S.; {Hinder}, I.; {Ohme}, F.;  et~al.
\newblock {Science with the space-based interferometer eLISA: Supermassive
  black hole binaries}.
\newblock {\em Phys. Rev. D Part. Fields Gravit. Cosmol. } {\bf 2016}, {\em 93},~024003.
\newblock {\url{https://doi.org/10.1103/PhysRevD.93.024003}}.

\bibitem[{Amaro-Seoane} et~al.(2017){Amaro-Seoane}, {Audley}, {Babak}, {Baker},
  {Barausse}, {Bender}, {Berti}, {Binetruy}, {Born}, {Bortoluzzi}, {Camp},
  {Caprini}, {Cardoso}, {Colpi}, {Conklin}, {Cornish}, {Cutler}, {Danzmann},
  {Dolesi}, {Ferraioli}, {Ferroni}, {Fitzsimons}, {Gair}, {Gesa Bote},
  {Giardini}, {Gibert}, {Grimani}, {Halloin}, {Heinzel}, {Hertog}, {Hewitson},
  {Holley-Bockelmann}, {Hollington}, {Hueller}, {Inchauspe}, {Jetzer},
  {Karnesis}, {Killow}, {Klein}, {Klipstein}, {Korsakova}, {Larson}, {Livas},
  {Lloro}, {Man}, {Mance}, {Martino}, {Mateos}, {McKenzie}, {McWilliams},
  {Miller}, {Mueller}, {Nardini}, {Nelemans}, {Nofrarias}, {Petiteau},
  {Pivato}, {Plagnol}, {Porter}, {Reiche}, {Robertson}, {Robertson}, {Rossi},
  {Russano}, {Schutz}, {Sesana}, {Shoemaker}, {Slutsky}, {Sopuerta}, {Sumner},
  {Tamanini}, {Thorpe}, {Troebs}, {Vallisneri}, {Vecchio}, {Vetrugno},
  {Vitale}, {Volonteri}, {Wanner}, {Ward}, {Wass}, {Weber}, {Ziemer}, and
  {Zweifel}]{amaro2017}
{Amaro-Seoane}, P.; {Audley}, H.; {Babak}, S.; {Baker}, J.; {Barausse}, E.;
  {Bender}, P.; {Berti}, E.; {Binetruy}, P.; {Born}, M.; {Bortoluzzi}, D.;
  et~al.
\newblock {Laser Interferometer Space Antenna}.
\newblock {\em arXiv } {\bf 2017}, arXiv:1702.00786.
\newblock {\url{https://doi.org/10.48550/arXiv.1702.00786}}.

\bibitem[{Mo{\'s}cibrodzka} et~al.(2011){Mo{\'s}cibrodzka}, {Gammie},
  {Dolence}, and {Shiokawa}]{Monika2011}
{Mo{\'s}cibrodzka}, M.; {Gammie}, C.F.; {Dolence}, J.C.; {Shiokawa}, H.
\newblock {Pair Production in Low-luminosity Galactic Nuclei}.
\newblock {\em Astrophys. J. } {\bf 2011}, {\em 735},~9.
\newblock {\url{https://doi.org/10.1088/0004-637X/735/1/9}}.

\bibitem[{Wong} et~al.(2021){Wong}, {Ryan}, and {Gammie}]{wong2021}
{Wong}, G.N.; {Ryan}, B.R.; {Gammie}, C.F.
\newblock {Pair Drizzle around Sub-Eddington Supermassive Black Holes}.
\newblock {\em Astrophys. J. } {\bf 2021}, {\em 907},~73.
\newblock {\url{https://doi.org/10.3847/1538-4357/abd0f9}}.

\bibitem[{Kimura} et~al.(2022){Kimura}, {Toma}, {Noda}, and {Hada}]{kimura2022}
{Kimura}, S.S.; {Toma}, K.; {Noda}, H.; {Hada}, K.
\newblock {Magnetic Reconnection in Black Hole Magnetospheres: Lepton Loading
  into Jets, Superluminal Radio Blobs, and Multiwavelength Flares}.
\newblock {\em Astrophys. J. Lett.} {\bf 2022}, {\em 937},~L34.
\newblock {\url{https://doi.org/10.3847/2041-8213/ac8d5a}}.

\bibitem[{Globus} and {Blandford}(2023)]{globus2023}
{Globus}, N.; {Blandford}, R.
\newblock {Ultra High Energy Cosmic Ray Source Models: Successes, Challenges
  and General Predictions}.
\newblock \emph{EPJ Web Conf.} \textbf{2023}, \emph{283}, 04001.
\newblock {\url{https://doi.org/10.1051/epjconf/202328304001}}.

\bibitem[{M{\'e}sz{\'a}ros}(2017)]{meszaros2017}
{M{\'e}sz{\'a}ros}, P.
\newblock {Astrophysical Sources of High-Energy Neutrinos in the IceCube Era}.
\newblock {\em Annu. Rev. Nucl. Part. Sci.} {\bf 2017}, {\em
  67},~45--67.
\newblock {\url{https://doi.org/10.1146/annurev-nucl-101916-123304}}.

\bibitem[{IceCube Collaboration} et~al.(2018){IceCube Collaboration},
  {Aartsen}, {Ackermann}, {Adams}, {Aguilar}, {Ahlers}, {Ahrens}, {Samarai},
  {Altmann}, {Andeen}, {Anderson}, {Ansseau}, {Anton}, {Arg{\"u}elles},
  {Arsioli}, {Auffenberg}, {Axani}, {Bagherpour}, {Bai}, {Barron}, {Barwick},
  {Baum}, {Bay}, {Beatty}, {Becker Tjus}, {Becker}, {BenZvi}, {Berley},
  {Bernardini}, {Besson}, {Binder}, {Bindig}, {Blaufuss}, {Blot}, {Bohm},
  {B{\"o}rner}, {Bos}, {B{\"o}ser}, {Botner}, {Bourbeau}, {Bourbeau},
  {Bradascio}, {Braun}, {Brenzke}, {Bretz}, {Bron}, {Brostean-Kaiser},
  {Burgman}, {Busse}, {Carver}, {Cheung}, {Chirkin}, {Christov}, {Clark},
  {Classen}, {Coenders}, {Collin}, {Conrad}, {Coppin}, {Correa}, {Cowen},
  {Cross}, {Dave}, {Day}, {de Andr{\'e}}, {De Clercq}, {DeLaunay}, {Dembinski},
  {DeRidder}, {Desiati}, {de Vries}, {de Wasseige}, {de With}, {DeYoung},
  {D{\'\i}az-V{\'e}lez}, {di Lorenzo}, {Dujmovic}, {Dumm}, {Dunkman}, {Dvorak},
  {Eberhardt}, {Ehrhardt}, {Eichmann}, {Eller}, {Evenson}, {Fahey}, {Fazely},
  {Felde}, {Filimonov}, {Finley}, {Flis}, {Franckowiak}, {Friedman}, {Fritz},
  {Gaisser}, {Gallagher}, {Gerhardt}, {Ghorbani}, {Giommi}, {Glauch},
  {Gl{\"u}senkamp}, {Goldschmidt}, {Gonzalez}, {Grant}, {Griffith}, {Haack},
  {Hallgren}, {Halzen}, {Hanson}, {Hebecker}, {Heereman}, {Helbing},
  {Hellauer}, {Hickford}, {Hignight}, {Hill}, {Hoffman}, {Hoffmann}, {Hoinka},
  {Hokanson-Fasig}, {Hoshina}, {Huang}, {Huber}, {Hultqvist}, {H{\"u}nnefeld},
  {Hussain}, {In}, {Iovine}, {Ishihara}, {Jacobi}, {Japaridze}, {Jeong},
  {Jero}, {Jones}, {Kalaczynski}, {Kang}, {Kappes}, {Kappesser}, {Karg},
  {Karle}, {Katz}, {Kauer}, {Keivani}, {Kelley}, {Kheirandish}, {Kim}, {Kim},
  {Kintscher}, {Kiryluk}, {Kittler}, {Klein}, {Koirala}, {Kolanoski},
  {K{\"o}pke}, {Kopper}, {Kopper}, {Koschinsky}, {Koskinen}, {Kowalski},
  {Krammer}, {Krings}, {Kroll}, {Kr{\"u}ckl}, {Kunwar}, {Kurahashi},
  {Kuwabara}, {Kyriacou}, {Labare}, {Lanfranchi}, {Larson}, {Lauber},
  {Leonard}, {Lesiak-Bzdak}, {Leuermann}, {Liu}, {Lozano Mariscal}, {Lu},
  {L{\"u}nemann}, {Luszczak}, {Madsen}, {Maggi}, {Mahn}, {Mancina}, {Maruyama},
  {Mase}, {Maunu}, {Meagher}, {Medici}, {Meier}, {Menne}, {Merino}, {Meures},
  {Miarecki}, {Micallef}, {Moment{\'e}}, {Montaruli}, {Moore}, {Morse},
  {Moulai}, and {Nahnhauer}]{icecube2018}
{IceCube Collaboration}.; {Aartsen}, M.G.; {Ackermann}, M.; {Adams}, J.;
  {Aguilar}, J.A.; {Ahlers}, M.; {Ahrens}, M.; {Samarai}, I.A.; {Altmann}, D.;
  {Andeen}, K.;  et~al.
\newblock {Neutrino emission from the direction of the blazar TXS 0506+056
  prior to the IceCube-170922A alert}.
\newblock {\em Science} {\bf 2018}, {\em 361},~147--151.
\newblock {\url{https://doi.org/10.1126/science.aat2890}}.

\bibitem[{Abbasi} et~al.(2024){Abbasi}, {Ackermann}, {Adams}, {Agarwalla},
  {Aguilar}, {Ahlers}, {Alameddine}, {Amin}, {Andeen}, {Arg{\"u}elles},
  {Ashida}, {Athanasiadou}, {Ausborm}, {Axani}, {Bai}, {Balagopal},
  {Baricevic}, {Barwick}, {Bash}, {Basu}, {Bay}, {Beatty}, {Becker Tjus},
  {Beise}, {Bellenghi}, {Benning}, {BenZvi}, {Berley}, {Bernardini}, {Besson},
  {Blaufuss}, {Bloom}, {Blot}, {Bontempo}, {Book Motzkin}, {Boscolo Meneguolo},
  {B{\"o}ser}, {Botner}, {B{\"o}ttcher}, {Braun}, {Brinson}, {Brostean-Kaiser},
  {Brusa}, {Burley}, {Butterfield}, {Campana}, {Caracas}, {Carloni}, {Carpio},
  {Chattopadhyay}, {Chau}, {Chen}, {Chirkin}, {Choi}, {Clark}, {Coleman},
  {Collin}, {Connolly}, {Conrad}, {Corley}, {Cowen}, {Dave}, {De Clercq},
  {DeLaunay}, {Delgado}, {Deng}, {Desai}, {Desiati}, {de Vries}, {de Wasseige},
  {DeYoung}, {Diaz}, {D{\'\i}az-V{\'e}lez}, {Dierichs}, {Dittmer}, {Domi},
  {Draper}, {Dujmovic}, {Durnford}, {Dutta}, {DuVernois}, {Ehrhardt},
  {Eidenschink}, {Eimer}, {Eller}, {Ellinger}, {El Mentawi}, {Els{\"a}sser},
  {Engel}, {Erpenbeck}, {Evans}, {Evenson}, {Fan}, {Fang}, {Farrag}, {Fazely},
  {Fedynitch}, {Feigl}, {Fiedlschuster}, {Finley}, {Fischer}, {Fox},
  {Franckowiak}, {Fukami}, {F{\"u}rst}, {Gallagher}, {Ganster}, {Garcia},
  {Garcia}, {Garg}, {Genton}, {Gerhardt}, {Ghadimi}, {Girard-Carillo},
  {Glaser}, {Gl{\"u}senkamp}, {Gonzalez}, {Goswami}, {Granados}, {Grant},
  {Gray}, {Gries}, {Griffin}, {Griswold}, {Groth}, {Guevel}, {G{\"u}nther},
  {Gutjahr}, {Ha}, {Haack}, {Hallgren}, {Halve}, {Halzen}, {Hamdaoui}, {Minh},
  {Handt}, {Hanson}, {Hardin}, {Harnisch}, {Hatch}, {Haungs},
  {H{\"a}u{\ss}ler}, {Helbing}, {Hellrung}, {Hermannsgabner}, {Heuermann},
  {Heyer}, {Hickford}, {Hidvegi}, {Hill}, {Hill}, {Hoffman}, {Hori}, {Hoshina},
  {Hostert}, {Hou}, {Huber}, {Hultqvist}, {H{\"u}nnefeld}, {Hussain}, {Hymon},
  {Ishihara}, {Iwakiri}, {Jacquart}, {Jain}, {Janik}, {Jansson}, {Japaridze},
  {Jeong}, {Jin}, {Jones}, {Kamp}, {Kang}, {Kang}, {Kang}, {Kappes},
  {Kappesser}, {Kardum}, {Karg}, {Karl}, {Karle}, {Katil}, {Katz}, {Kauer},
  {Kelley}, {Khanal}, {Khatee Zathul}, {Kheirandish}, {Kiryluk}, {Klein},
  {Kochocki}, {Koirala}, {Kolanoski}, {Kontrimas}, {K{\"o}pke}, {Kopper},
  {Koskinen}, {Koundal}, {Kovacevich}, and {Kowalski}]{abb2024}
{Abbasi}, R.; {Ackermann}, M.; {Adams}, J.; {Agarwalla}, S.K.; {Aguilar}, J.A.;
  {Ahlers}, M.; {Alameddine}, J.M.; {Amin}, N.M.; {Andeen}, K.;
  {Arg{\"u}elles}, C.;  et~al.
\newblock {Probing the Connection between IceCube Neutrinos and MOJAVE AGN}.
\newblock {\em Astrophys. J. } {\bf 2024}, {\em 973},~97.
\newblock {\url{https://doi.org/10.3847/1538-4357/ad643d}}.

\bibitem[{Dermer} and {Menon}(2009)]{dermer2009}
{Dermer}, C.D.; {Menon}, G.
\newblock {\em {High Energy Radiation from Black Holes: Gamma Rays, Cosmic
  Rays, and Neutrinos}}; Princeton University Press: Princeton, NJ,  USA, 2009.

\bibitem[{Rybicki} and {Lightman}(1980)]{rybicki1986}
{Rybicki}, G.B.; {Lightman}, A.P.
\newblock {\em {Radiative Processes in Astrophysics}}; John Wiley \& Sons Inc: {Hoboken, NJ, USA,} 
  1980.

\bibitem[{Anantua} et~al.(2020){Anantua}, {Emami}, {Loeb}, and
  {Chael}]{ana2020}
{Anantua}, R.; {Emami}, R.; {Loeb}, A.; {Chael}, A.
\newblock {Determining the Composition of Relativistic Jets from Polarization
  Maps}.
\newblock {\em Astrophys. J. } {\bf 2020}, {\em 896},~30.
\newblock {\url{https://doi.org/10.3847/1538-4357/ab9103}}.

\bibitem[{Mahadevan}(1997)]{mahadevan1997}
{Mahadevan}, R.
\newblock {Scaling Laws for Advection-dominated Flows: Applications to
  Low-Luminosity Galactic Nuclei}.
\newblock {\em Astrophys. J. } {\bf 1997}, {\em 477},~585--601.
\newblock {\url{https://doi.org/10.1086/303727}}.

\bibitem[{Jakubassa-Amundsen} and
  {Mangiarotti}(2019)]{ff_cross_2019PhRvA.100c2703J}
{Jakubassa-Amundsen}, D.H.; {Mangiarotti}, A.
\newblock {Accuracy of analytical theories for relativistic bremsstrahlung}.
\newblock {\em Phys. Rev. A} {\bf 2019}, {\em 100},~032703.
\newblock {\url{https://doi.org/10.1103/PhysRevA.100.032703}}.

\bibitem[{Owen} et~al.(2023){Owen}, {Wu}, {Inoue}, {Yang}, and
  {Mitchell}]{owen2023}
{Owen}, E.R.; {Wu}, K.; {Inoue}, Y.; {Yang}, H.Y.K.; {Mitchell}, A.M.W.
\newblock {Cosmic Ray Processes in Galactic Ecosystems}.
\newblock {\em Galaxies} {\bf 2023}, {\em 11},~86.
\newblock {\url{https://doi.org/10.3390/galaxies11040086}}.

\bibitem[{Jacobsen} et~al.(2015){Jacobsen}, {Wu}, {On}, and
  {Saxton}]{jacobsen2015}
{Jacobsen}, I.B.; {Wu}, K.; {On}, A.Y.L.; {Saxton}, C.J.
\newblock {High-energy neutrino fluxes from AGN populations inferred from X-ray
  surveys}.
\newblock {\em Mon. Not. R. Astron. Soc.} {\bf 2015}, {\em 451},~3649--3663.
\newblock {\url{https://doi.org/10.1093/mnras/stv1196}}.

\bibitem[{Murase}(2017)]{murase2017}
{Murase}, K.
\newblock {Active Galactic Nuclei as High-Energy Neutrino Sources}. In {\em
  Neutrino Astronomy: Current Status, Future Prospects}; {Gaisser}, T.,
  {Karle}, A., Eds.;  {World Scientific Publishing: Singapore; Hackensack, NJ, USA,} 
 2017; pp. 15--31.
\newblock {\url{https://doi.org/10.1142/9789814759410_0002}}.

\bibitem[{Odell} and {Gooding}(1986)]{Odell1986}
{Odell}, A.W.; {Gooding}, R.H.
\newblock {Procedures for Solving Kepler's Equation}.
\newblock {\em Celest. Mech.} {\bf 1986}, {\em 38},~307--334.
\newblock {\url{https://doi.org/10.1007/BF01238923}}.

\bibitem[{Murray} and {Dermott}(1999)]{murray1999}
{Murray}, C.D.; {Dermott}, S.F.
\newblock {\em {Solar System Dynamics}};  {Cambridge University Press: Cambridge, UK,} 1999.
\newblock {\url{https://doi.org/10.1017/CBO9781139174817}}.

\bibitem[{Kafexhiu} et~al.(2014){Kafexhiu}, {Aharonian}, {Taylor}, and
  {Vila}]{kafexhiu2014}
{Kafexhiu}, E.; {Aharonian}, F.; {Taylor}, A.M.; {Vila}, G.S.
\newblock {Parametrization of gamma-ray production cross sections for p p
  interactions in a broad proton energy range from the kinematic threshold to
  PeV energies}.
\newblock {\em Phys. Rev. D Part. Fields Gravit. Cosmol. } {\bf 2014}, {\em 90},~123014.
\newblock {\url{https://doi.org/10.1103/PhysRevD.90.123014}}.

\bibitem[{Shakura} and {Sunyaev}(1973)]{Shakura1973}
{Shakura}, N.I.; {Sunyaev}, R.A.
\newblock {Black holes in binary systems. Observational appearance.}
\newblock {\em Astron. Astrophys. } {\bf 1973}, {\em 24},~337--355.

\bibitem[{Novikov} and {Thorne}(1973)]{nov73}
{Novikov}, I.; {Thorne}, K.
\newblock {Black Holes}.
\newblock {\em Les Astres Occlus}; Gordon Breach: New York, NY, USA, 1973; Volume
  52,~pp. 343--450.

\bibitem[{Kelner} et~al.(2006){Kelner}, {Aharonian}, and {Bugayov}]{kelner2006}
{Kelner}, S.R.; {Aharonian}, F.A.; {Bugayov}, V.V.
\newblock {Energy spectra of gamma rays, electrons, and neutrinos produced at
  proton-proton interactions in the very high energy regime}.
\newblock {\em Phys. Rev. D Part. Fields Gravit. Cosmol. } {\bf 2006}, {\em 74},~034018.
\newblock {\url{https://doi.org/10.1103/PhysRevD.74.034018}}.

\bibitem[{Bondi} and {Hoyle}(1944)]{bon44}
{Bondi}, H.; {Hoyle}, F.
\newblock {On the mechanism of accretion by stars}.
\newblock {\em Mon. Not. R. Astron. Soc.} {\bf 1944}, {\em 104},~273.
\newblock {\url{https://doi.org/10.1093/mnras/104.5.273}}.

\bibitem[{Bondi}(1952)]{bon52}
{Bondi}, H.
\newblock {On spherically symmetrical accretion}.
\newblock {\em Mon. Not. R. Astron. Soc.} {\bf 1952}, {\em 112},~195.
\newblock {\url{https://doi.org/10.1093/mnras/112.2.195}}.

\bibitem[{Pringle}(1981)]{pringle1981}
{Pringle}, J.E.
\newblock {Accretion discs in astrophysics}.
\newblock {\em Annu. Rev. Astron. Astrophys. } {\bf 1981}, {\em 19},~137--162.
\newblock {\url{https://doi.org/10.1146/annurev.aa.19.090181.001033}}.

\bibitem[{Tanaka}(2011)]{tanaka2011}
{Tanaka}, T.
\newblock {Exact time-dependent solutions for the thin accretion disc equation:
  boundary conditions at finite radius}.
\newblock {\em Mon. Not. R. Astron. Soc.} {\bf 2011}, {\em 410},~1007--1017.
\newblock {\url{https://doi.org/10.1111/j.1365-2966.2010.17496.x}}.

\bibitem[{Kato} et~al.(1998){Kato}, {Fukue}, and {Mineshige}]{kato1998}
{Kato}, S.; {Fukue}, J.; {Mineshige}, S.
\newblock {\em {Black-Hole Accretion Disks--Toward a New Paradigm}}; {Kyoto University Press: Kyoto, Japan,} 1998.

\bibitem[{Abramowicz} et~al.(1988){Abramowicz}, {Czerny}, {Lasota}, and
  {Szuszkiewicz}]{slim1988}
{Abramowicz}, M.A.; {Czerny}, B.; {Lasota}, J.P.; {Szuszkiewicz}, E.
\newblock {Slim Accretion Disks}.
\newblock {\em Astrophys. J. } {\bf 1988}, {\em 332},~646.
\newblock {\url{https://doi.org/10.1086/166683}}.

\bibitem[{S{\k{a}}dowski} et~al.(2014){S{\k{a}}dowski}, {Narayan}, {McKinney},
  and {Tchekhovskoy}]{skadowski2014}
{S{\k{a}}dowski}, A.; {Narayan}, R.; {McKinney}, J.C.; {Tchekhovskoy}, A.
\newblock {Numerical simulations of super-critical black hole accretion flows
  in general relativity}.
\newblock {\em Mon. Not. R. Astron. Soc.} {\bf 2014}, {\em 439},~503--520.
\newblock {\url{https://doi.org/10.1093/mnras/stt2479}}.

\bibitem[{Ricarte} et~al.(2023){Ricarte}, {Narayan}, and {Curd}]{ricarte2023}
{Ricarte}, A.; {Narayan}, R.; {Curd}, B.
\newblock {Recipes for Jet Feedback and Spin Evolution of Black Holes with
  Strongly Magnetized Super-Eddington Accretion Disks}.
\newblock {\em Astrophys. J. Lett.} {\bf 2023}, {\em 954},~L22.
\newblock {\url{https://doi.org/10.3847/2041-8213/aceda5}}.

\bibitem[{Curd} et~al.(2023){Curd}, {Emami}, {Anantua}, {Palumbo}, {Doeleman},
  and {Narayan}]{curd2023}
{Curd}, B.; {Emami}, R.; {Anantua}, R.; {Palumbo}, D.; {Doeleman}, S.;
  {Narayan}, R.
\newblock {Jets from SANE super-Eddington accretion discs: Morphology, spectra,
  and their potential as targets for ngEHT}.
\newblock {\em Mon. Not. R. Astron. Soc.} {\bf 2023}, {\em 519},~2812--2837.
\newblock {\url{https://doi.org/10.1093/mnras/stac3716}}.

\bibitem[{Esin} et~al.(1997){Esin}, {McClintock}, and {Narayan}]{esin1997}
{Esin}, A.A.; {McClintock}, J.E.; {Narayan}, R.
\newblock {Advection-Dominated Accretion and the Spectral States of Black Hole
  X-Ray Binaries: Application to Nova Muscae 1991}.
\newblock {\em Astrophys. J. } {\bf 1997}, {\em 489},~865--889.
\newblock {\url{https://doi.org/10.1086/304829}}.

\bibitem[{Remillard} and {McClintock}(2006)]{remillard2006}
{Remillard}, R.A.; {McClintock}, J.E.
\newblock {X-Ray Properties of Black-Hole Binaries}.
\newblock {\em Annu. Rev. Astron. Astrophys. } {\bf 2006}, {\em 44},~49--92.
\newblock {\url{https://doi.org/10.1146/annurev.astro.44.051905.092532}}.

\bibitem[{The IceCube Collaboration} et~al.(2015){The IceCube Collaboration},
  {Aartsen}, {Abraham}, {Ackermann}, {Adams}, {Aguilar}, {Ahlers}, {Ahrens},
  {Altmann}, {Anderson}, {Ansseau}, {Archinger}, {Arguelles}, {Arlen},
  {Auffenberg}, {Bai}, {Barwick}, {Baum}, {Bay}, {Beatty}, {Becker Tjus},
  {Becker}, {Beiser}, {BenZvi}, {Berghaus}, {Berley}, {Bernardini}, {Bernhard},
  {Besson}, {Binder}, {Bindig}, {Bissok}, {Blaufuss}, {Blumenthal}, {Boersma},
  {Bohm}, {B{\"o}rner}, {Bos}, {Bose}, {B{\"o}ser}, {Botner}, {Braun},
  {Brayeur}, {Bretz}, {Buzinsky}, {Casey}, {Casier}, {Cheung}, {Chirkin},
  {Christov}, {Clark}, {Classen}, {Coenders}, {Cowen}, {Cruz Silva},
  {Daughhetee}, {Davis}, {Day}, {de Andr{\'e}}, {De Clercq}, {del Pino
  Rosendo}, {Dembinski}, {De Ridder}, {Desiati}, {de Vries}, {de Wasseige}, {de
  With}, {DeYoung}, {Diaz-V{\'e}lez}, {di Lorenzo}, {Dumm}, {Dunkman}, {Eagan},
  {Eberhardt}, {Ehrhardt}, {Eichmann}, {Euler}, {Evenson}, {Fadiran}, {Fahey},
  {Fazely}, {Fedynitch}, {Feintzeig}, {Felde}, {Filimonov}, {Finley},
  {Fischer-Wasels}, {Flis}, {F{\"o}sig}, {Fuchs}, {Gaisser}, {Gaior},
  {Gallagher}, {Gerhardt}, {Ghorbani}, {Gier}, {Gladstone}, {Glagla},
  {Gl{\"u}senkamp}, {Goldschmidt}, {Golup}, {Gonzalez}, {G{\'o}ra}, {Grant},
  {Groh}, {Gro{\ss}}, {Ha}, {Haack}, {Haj Ismail}, {Hallgren}, {Halzen},
  {Hansmann}, {Hanson}, {Hebecker}, {Heereman}, {Helbing}, {Hellauer},
  {Hellwig}, {Hickford}, {Hignight}, {Hill}, {Hoffman}, {Hoffmann},
  {Holzapfel}, {Homeier}, {Hoshina}, {Huang}, {Huber}, {Huelsnitz}, {Hulth},
  {Hultqvist}, {In}, {Ishihara}, {Jacobi}, {Japaridze}, {Jero}, {Jurkovic},
  {Kaminsky}, {Kappes}, {Karg}, {Karle}, {Kauer}, {Keivani}, {Kelley}, {Kemp},
  {Kheirandish}, {Kiryluk}, {Kl{\"a}s}, {Klein}, {Kohnen}, {Koirala},
  {Kolanoski}, {Konietz}, {Koob}, {K{\"o}pke}, {Kopper}, {Kopper}, {Koskinen},
  {Kowalski}, {Krings}, {Kroll}, {Kroll}, {Kunnen}, {Kurahashi}, {Kuwabara},
  {Labare}, {Lanfranchi}, {Larson}, {Lesiak-Bzdak}, {Leuermann}, {Leuner},
  {Lu}, {L{\"u}nemann}, {Madsen}, {Maggi}, {Mahn}, {Maruyama}, {Mase}, {Matis},
  {Maunu}, {McNally}, {Meagher}, {Medici}, {Meli}, {Menne}, {Merino}, {Meures},
  {Miarecki}, {Middell}, {Middlemas}, {Mohrmann}, {Montaruli}, {Morse},
  {Nahnhauer}, {Naumann}, {Neer}, {Niederhausen}, {Nowicki}, {Nygren}, and
  {Obertacke}]{icecube2015}
{The IceCube Collaboration}; {Aartsen}, M.G.; {Abraham}, K.; {Ackermann}, M.;
  {Adams}, J.; {Aguilar}, J.A.; {Ahlers}, M.; {Ahrens}, M.; {Altmann}, D.;
  {Anderson}, T.;  et~al.
\newblock {The IceCube Neutrino Observatory---Contributions to ICRC 2015 Part
  I: Point Source Searches}.
\newblock {\em arXiv} {\bf 2015}, arXiv:1510.05222.
\newblock {\url{https://doi.org/10.48550/arXiv.1510.05222}}.

\bibitem[{Aartsen} et~al.(2017){Aartsen}, {Ackermann}, {Adams}, {Aguilar},
  {Ahlers}, {Ahrens}, {Altmann}, {Andeen}, {Anderson}, {Ansseau}, {Anton},
  {Archinger}, {Arg{\"u}elles}, {Auer}, {Auffenberg}, {Axani}, {Baccus}, {Bai},
  {Barnet}, {Barwick}, {Baum}, {Bay}, {Beattie}, {Beatty}, {Becker Tjus},
  {Becker}, {Bendfelt}, {BenZvi}, {Berley}, {Bernardini}, {Bernhard}, {Besson},
  {Binder}, {Bindig}, {Bissok}, {Blaufuss}, {Blot}, {Boersma}, {Bohm},
  {B{\"o}rner}, {Bos}, {Bose}, {B{\"o}ser}, {Botner}, {Bouchta}, {Braun},
  {Brayeur}, {Bretz}, {Bron}, {Burgman}, {Burreson}, {Carver}, {Casier},
  {Cheung}, {Chirkin}, {Christov}, {Clark}, {Classen}, {Coenders}, {Collin},
  {Conrad}, {Cowen}, {Cross}, {Day}, {Day}, {de Andr{\'e}}, {De Clercq}, {del
  Pino Rosendo}, {Dembinski}, {De Ridder}, {Descamps}, {Desiati}, {de Vries},
  {de Wasseige}, {de With}, {DeYoung}, {D{\'\i}az-V{\'e}lez}, {di Lorenzo},
  {Dujmovic}, {Dumm}, {Dunkman}, {Eberhardt}, {Edwards}, {Ehrhardt},
  {Eichmann}, {Eller}, {Euler}, {Evenson}, {Fahey}, {Fazely}, {Feintzeig},
  {Felde}, {Filimonov}, {Finley}, {Flis}, {F{\"o}sig}, {Franckowiak},
  {Fr{\`e}re}, {Friedman}, {Fuchs}, {Gaisser}, {Gallagher}, {Gerhardt},
  {Ghorbani}, {Giang}, {Gladstone}, {Glauch}, {Glowacki}, {Gl{\"u}senkamp},
  {Goldschmidt}, {Gonzalez}, {Grant}, {Griffith}, {Gustafsson}, {Haack},
  {Hallgren}, {Halzen}, {Hansen}, {Hansmann}, {Hanson}, {Haugen}, {Hebecker},
  {Heereman}, {Helbing}, {Hellauer}, {Heller}, {Hickford}, {Hignight}, {Hill},
  {Hoffman}, {Hoffmann}, {Hoshina}, {Huang}, {Huber}, {Hulth}, {Hultqvist},
  {In}, {Inaba}, {Ishihara}, {Jacobi}, {Jacobsen}, {Japaridze}, {Jeong},
  {Jero}, {Jones}, {Jones}, {Joseph}, {Kang}, {Kappes}, {Karg}, {Karle},
  {Katz}, {Kauer}, {Keivani}, {Kelley}, {Kemp}, {Kheirandish}, {Kim}, {Kim},
  {Kintscher}, {Kiryluk}, {Kitamura}, {Kittler}, {Klein}, {Kleinfelder},
  {Kleist}, {Kohnen}, {Koirala}, {Kolanoski}, {Konietz}, {K{\"o}pke}, {Kopper},
  {Kopper}, {Koskinen}, {Kowalski}, {Krasberg}, {Krings}, {Kroll},
  {Kr{\"u}ckl}, {Kr{\"u}ger}, {Kunnen}, {Kunwar}, {Kurahashi}, {Kuwabara},
  {Labare}, {Laihem}, {Landsman}, {Lanfranchi}, {Larson}, {Lauber}, {Laundrie},
  {Lennarz}, {Leich}, {Lesiak-Bzdak}, {Leuermann}, {Lu}, {Ludwig},
  {L{\"u}nemann}, {Mackenzie}, and {Madsen}]{icecube2017}
{Aartsen}, M.G.; {Ackermann}, M.; {Adams}, J.; {Aguilar}, J.A.; {Ahlers}, M.;
  {Ahrens}, M.; {Altmann}, D.; {Andeen}, K.; {Anderson}, T.; {Ansseau}, I.;
  et~al.
\newblock {The IceCube Neutrino Observatory: Instrumentation and online
  systems}.
\newblock {\em J. Instrum.} {\bf 2017}, {\em 12},~P03012.
\newblock {\url{https://doi.org/10.1088/1748-0221/12/03/P03012}}.

\bibitem[{Sinopoulou} et~al.(2021){Sinopoulou}, {Coniglione}, {Muller},
  {Tzamariudaki}, {Tzamariudaki}, and {KM3NeT collaboration}]{KM3NET2021}
{Sinopoulou}, A.; {Coniglione}, R.; {Muller}, R.; {Tzamariudaki}, E.;
  {Tzamariudaki}, E.; {KM3NeT Collaboration}.
\newblock {Atmospheric neutrinos with the first detection units of
  KM3NeT/ARCA}.
\newblock {\em J. Instrum.} {\bf 2021}, {\em 16},~C11015.
\newblock {\url{https://doi.org/10.1088/1748-0221/16/11/C11015}}.

\bibitem[{The KM3NeT Collaboration} et~al.(2025){The KM3NeT Collaboration},
  {Aiello}, {Albert}, {Alhebsi}, {Alshamsi}, {Alves Garre}, {Ambrosone},
  {Ameli}, {Andre}, {Anghinolfi}, {Aphecetche}, {Ardid}, {Ardid},
  {Arg{\"u}elles}, {Atmani}, {Aublin}, {Badaracco}, {Bailly-Salins},
  {Barda{\v{c}}ov{\'a}}, {Baret}, {Bariego-Quintana}, {Becherini}, {Bendahman},
  {Benfenati Gualandi}, {Benhassi}, {Bennani}, {Benoit}, {Berbee}, {Bertin},
  {Biagi}, {Boettcher}, {Bonanno}, {Bouasla}, {Boumaaza}, {Bouta}, {Bouwhuis},
  {Bozza}, {Bozza}, {Br{\^a}nza{\c{s}}}, {Bretaudeau}, {Breuhaus}, {Bruijn},
  {Brunner}, {Bruno}, {Buis}, {Buompane}, {Buson}, {Busto}, {Caiffi}, {Calvo},
  {Capone}, {Carenini}, {Carretero}, {Cartraud}, {Castaldi}, {Cecchini},
  {Celli}, {Cerisy}, {Chabab}, {Chen}, {Cherubini}, {Chiarusi}, {Circella},
  {Cocimano}, {Coelho}, {Coleiro}, {Colonges}, {Condorelli}, {Coniglione},
  {Coyle}, {Creusot}, {Cuttone}, {D'Amico}, {Dallier}, {De Benedittis}, {De
  Martino}, {De Wasseige}, {Decoene}, {Del Rosso}, {Di Mauro}, {Di Palma},
  {Diaz}, {Diego-Tortosa}, {Distefano}, {Domi}, {Donzaud}, {Dornic},
  {Drakopoulou}, {Drouhin}, {Ducoin}, {Dvornick{\'y}}, {Eberl}, {Eckerov{\'a}},
  {Eddymaoui}, {van Eeden}, {Eff}, {van Eijk}, {El Bojaddaini}, {El Hedri},
  {Ellajosyula}, {Enzenh{\"o}fer}, {Ferrara}, {Filipovi{\'c}}, {Filippini},
  {Franciotti}, {Fusco}, {Gagliardini}, {Gal}, {Garc{\'\i}a M{\'e}ndez},
  {Garcia Soto}, {Gatius Oliver}, {Gei{\ss}elbrecht}, {Genton}, {Ghaddari},
  {Gialanella}, {Gibson}, {Giorgio}, {Goos}, {Goswami}, {Gozzini}, {Gracia},
  {Graf}, {Guidi}, {Guillon}, {Guti{\'e}rrez}, {Haack}, {van Haren},
  {Heijboer}, {Hennig}, {Henry}, {Hern{\'a}ndez-Rey}, {Idrissi Ibnsalih},
  {Ilioni}, {Illuminati}, {Joly}, {de Jong}, {de Jong}, {Jung},
  {Kalaczy{\'n}ski}, {Kalekin}, {Kamp}, {Katz}, {Kistauri}, {Kopper},
  {Kouchner}, {Kovalev}, {Kueviakoe}, {Kulikovskiy}, {Kvatadze}, {Labalme},
  {Lahmann}, {Lamoureux}, {Lancelin}, {Larosa}, {Lastoria}, {Lazar}, {Lazo},
  {Le Stum}, {Lehaut}, {Lemaitre}, {Leonora}, {Lessing}, {Levi}, {Lincetto},
  {Lindsey Clark}, {Longhitano}, {Lumb}, {Magnani}, {Majumdar}, {Malerba},
  {Mamedov}, {Manfreda}, {Marconi}, {Margiotta}, {Marinelli}, {Markou},
  {Martin}, {Marzaioli}, {Mastrodicasa}, {Mastroianni}, {Mauro}, {Miele},
  {Migliozzi}, {Migneco}, {Mitsou}, {Mollo}, {Mongelli}, {Morales-Gallegos},
  {Moussa}, {Mozun Mateo}, {Muller}, {Musone}, {Musumeci}, {Navas},
  {Nayerhoda}, {Nicolau}, {Nkosi}, {Fearraigh}, {Oliviero}, and
  {Orlando}]{KM3NET2025}
{The KM3NeT Collaboration}; {Aiello}, S.; {Albert}, A.; {Alhebsi}, A.R.;
  {Alshamsi}, M.; {Alves Garre}, S.; {Ambrosone}, A.; {Ameli}, F.; {Andre}, M.;
  {Anghinolfi}, M.;  et~al.
\newblock {Observation of an ultra-high-energy cosmic neutrino with KM3NeT}.
\newblock {\em Nature } {\bf 2025}, {\em 638},~376--382.
\newblock {\url{https://doi.org/10.1038/s41586-024-08543-1}}.

\bibitem[{Baikal-GVD Collaboration} et~al.(2018){Baikal-GVD Collaboration},
  {:}, {Avrorin}, {Avrorin}, {Aynutdinov}, {Bannash}, {Belolaptikov},
  {Brudanin}, {Budnev}, {Doroshenko}, {Domogatsky}, {Dvornick{\'y}}, {Dyachok},
  {Dzhilkibaev}, {Fajt}, {Fialkovsky}, {Gafarov}, {Golubkov}, {Gres}, {Honz},
  {Kebkal}, {Kebkal}, {Khramov}, {Kolbin}, {Konischev}, {Korobchenko},
  {Koshechkin}, {Kozhin}, {Kulepov}, {Kuleshov}, {Milenin}, {Mirgazov},
  {Osipova}, {Panfilov}, {Pan'kov}, {Petukhov}, {Pliskovsky}, {Rozanov},
  {Rjabov}, {Rushay}, {Safronov}, {Simkovic}, {Shoibonov}, {Solovjev},
  {Sorokovikov}, {Shelepov}, {Suvorova}, {Shtekl}, {Tabolenko}, {Tarashansky},
  {Yakovlev}, {Zagorodnikov}, and {Zurbanov}]{baikal2018}
{Baikal-GVD Collaboration}; {Avrorin}, A.D.; {Avrorin}, A.V.;
  {Aynutdinov}, V.M.; {Bannash}, R.; {Belolaptikov}, I.A.; {Brudanin}, V.B.;
  {Budnev}, N.M.; {Doroshenko}, A.A.;  Domogatsky, G.V.; et~al.
\newblock {Baikal-GVD: Status and prospects}.
\newblock {\em arXiv } {\bf 2018},  arXiv:1808.10353.
\newblock {\url{https://doi.org/10.48550/arXiv.1808.10353}}.

\bibitem[{Schr{\"o}der}(2023)]{icecube-gen2}
{Schr{\"o}der}, F.G.
\newblock {Design and Expected Performance of the IceCube-Gen2 Surface Array
  and its Radio Component (ARENA2022)}.
\newblock {\em arXiv } {\bf 2023}, arXiv:2306.05900.
\newblock {\url{https://doi.org/10.48550/arXiv.2306.05900}}.

\bibitem[{Ishihara}(2019)]{pone2019}
{Ishihara}, A.
\newblock {The IceCube Upgrade---Design and Science Goals}.
\newblock In Proceedings of the 36th International Cosmic Ray Conference
  (ICRC2019),  {Madison, WI, USA, 24 July--1 August 2019;} 
 Volume~36, p.
  1031.
\newblock {\url{https://doi.org/10.22323/1.358.01031}}.

\bibitem[{Agostini} et~al.(2020){Agostini}, {B{\"o}hmer}, {Bosma}, {Clark},
  {Danninger}, {Fruck}, {Gernh{\"a}user}, {G{\"a}rtner}, {Grant}, {Henningsen},
  {Holzapfel}, {Huber}, {Jenkyns}, {Krauss}, {Krings}, {Kopper},
  {Leism{\"u}ller}, {Leys}, {Macoun}, {Meighen-Berger}, {Michel}, {Moore},
  {Morley}, {Padovani}, {Papp}, {Pirenne}, {Qiu}, {Rea}, {Resconi}, {Round},
  {Ruskey}, {Spannfellner}, {Traxler}, {Turcati}, and {Yanez}]{pone2020}
{Agostini}, M.; {B{\"o}hmer}, M.; {Bosma}, J.; {Clark}, K.; {Danninger}, M.;
  {Fruck}, C.; {Gernh{\"a}user}, R.; {G{\"a}rtner}, A.; {Grant}, D.;
  {Henningsen}, F.;  et~al.
\newblock {The Pacific Ocean Neutrino Experiment}.
\newblock {\em Nat. Astron.} {\bf 2020}, {\em 4},~913--915.
\newblock {\url{https://doi.org/10.1038/s41550-020-1182-4}}.

\bibitem[{Abi} et~al.(2020){Abi}, {Acciarri}, {Acero}, {Adamov}, {Adams},
  {Adinolfi}, {Ahmad}, {Ahmed}, {Alion}, {Alonso Monsalve}, {Alt}, {Anderson},
  {Andreopoulos}, {Andrews}, {Andrianala}, {Andringa}, {Ankowski}, {Antonova},
  {Antusch}, {Aranda-Fernandez}, {Ariga}, {Arnold}, {Arroyave}, {Asaadi},
  {Aurisano}, {Aushev}, {Autiero}, {Azfar}, {Back}, {Back}, {Backhouse},
  {Baesso}, {Bagby}, {Bajou}, {Balasubramanian}, {Baldi}, {Bambah}, {Barao},
  {Barenboim}, {Barker}, {Barkhouse}, {Barnes}, {Barr}, {Barranco Monarca},
  {Barros}, {Barrow}, {Bashyal}, {Basque}, {Bay}, {Bazo Alba}, {Beacom},
  {Bechetoille}, {Behera}, {Bellantoni}, {Bellettini}, {Bellini},
  {Beltramello}, {Belver}, {Benekos}, {Bento Neves}, {Berger}, {Berkman},
  {Bernardini}, {Berner}, {Berns}, {Bertolucci}, {Betancourt}, {Bezawada},
  {Bhattacharjee}, {Bhuyan}, {Biagi}, {Bian}, {Biassoni}, {Biery}, {Bilki},
  {Bishai}, {Bitadze}, {Blake}, {Blanco Siffert}, {Blaszczyk}, {Blazey},
  {Blucher}, {Boissevain}, {Bolognesi}, {Bolton}, {Bonesini}, {Bongrand},
  {Bonini}, {Booth}, {Booth}, {Bordoni}, {Borkum}, {Boschi}, {Bostan}, {Bour},
  {Boyd}, {Boyden}, {Bracinik}, {Braga}, {Brailsford}, {Brandt}, {Bremer},
  {Brew}, {Brianne}, {Brice}, {Brizzolari}, {Bromberg}, {Brooijmans}, {Brooke},
  {Bross}, {Brunetti}, {Buchanan}, {Budd}, {Caiulo}, {Calafiura}, {Calcutt},
  {Calin}, {Calvez}, {Calvo}, {Camilleri}, {Caminata}, {Campanelli},
  {Caratelli}, {Carini}, {Carlus}, {Carniti}, {Caro Terrazas}, {Carranza},
  {Castillo}, {Castromonte}, {Cattadori}, {Cavalier}, {Cavanna}, {Centro},
  {Cerati}, {Cervelli}, {Cervera Villanueva}, {Chalifour}, {Chang},
  {Chardonnet}, {Chatterjee}, {Chattopadhyay}, {Chaves}, {Chen}, {Chen},
  {Chen}, {Cherdack}, {Chi}, {Childress}, {Chiriacescu}, {Cho}, {Choubey},
  {Christensen}, {Christian}, {Christodoulou}, {Church}, {Clarke}, {Coan},
  {Cocco}, {Coelho}, {Conley}, {Conrad}, {Convery}, {Corwin}, {Cotte},
  {Cremaldi}, {Cremonesi}, {Crespo-Anad{\'o}n}, {Cristaldo}, {Cross}, {Cuesta},
  {Cui}, {Cussans}, {Dabrowski}, {Da Motta}, {Da Silva Peres}, {David},
  {Davies}, {Davini}, {Dawson}, {De}, {De Almeida}, {Debbins}, {De Bonis},
  {Decowski}, {De Gouvea}, {De Holanda}, {De Icaza Astiz}, {Deisting}, {De
  Jong}, {Delbart}, {Delepine}, {Delgado}, {Dell'Acqua}, {De Lurgio}, {De Mello
  Neto}, {DeMuth}, {Dennis}, {Densham}, and {Deptuch}]{dune2020}
{Abi}, B.; {Acciarri}, R.; {Acero}, M.A.; {Adamov}, G.; {Adams}, D.;
  {Adinolfi}, M.; {Ahmad}, Z.; {Ahmed}, J.; {Alion}, T.; {Alonso Monsalve}, S.;
   et~al.
\newblock {Volume I. Introduction to DUNE}.
\newblock {\em J. Instrum.} {\bf 2020}, {\em 15},~T08008.
\newblock {\url{https://doi.org/10.1088/1748-0221/15/08/T08008}}.

\bibitem[{Hyper-Kamiokande Proto-Collaboration} et~al.(2018){Hyper-Kamiokande
  Proto-Collaboration}, {:}, {Abe}, {Abe}, {Aihara}, {Aimi}, {Akutsu},
  {Andreopoulos}, {Anghel}, {Anthony}, {Antonova}, {Ashida}, {Aushev}, {Barbi},
  {Barker}, {Barr}, {Beltrame}, {Berardi}, {Bergevin}, {Berkman}, {Berns},
  {Berry}, {Bhadra}, {Bravo-Bergu{\~n}o}, {Blaszczyk}, {Blondel}, {Bolognesi},
  {Boyd}, {Bravar}, {Bronner}, {Buizza Avanzini}, {Cafagna}, {Cole}, {Calland},
  {Cao}, {Cartwright}, {Catanesi}, {Checchia}, {Chen-Wishart}, {Choi}, {Choi},
  {Coleman}, {Collazuol}, {Cowan}, {Cremonesi}, {Dealtry}, {De Rosa},
  {Densham}, {Dewhurst}, {Drakopoulou}, {Di Lodovico}, {Drapier}, {Dumarchez},
  {Dunne}, {Dziewiecki}, {Emery}, {Esmaili}, {Evangelisti},
  {Fernandez-Martinez}, {Feusels}, {Finch}, {Fiorentini}, {Fiorillo}, {Fitton},
  {Frankiewicz}, {Friend}, {Fujii}, {Fukuda}, {Fukuda}, {Ganezer}, {Giganti},
  {Gonin}, {Grant}, {Gumplinger}, {Hadley}, {Hartfiel}, {Hartz}, {Hayato},
  {Hayrapetyan}, {Hill}, {Hirota}, {Horiuchi}, {Ichikawa}, {Iijima}, {Ikeda},
  {Imber}, {Inoue}, {Insler}, {Intonti}, {Ioannisian}, {Ishida}, {Ishino},
  {Ishitsuka}, {Itow}, {Iwamoto}, {Izmaylov}, {Jamieson}, {Jang}, {Jang},
  {Jeon}, {Jiang}, {Jonsson}, {Joo}, {Kaboth}, {Kachulis}, {Kajita}, {Kameda},
  {Kataoka}, {Katori}, {Kayrapetyan}, {Kearns}, {Khabibullin}, {Khotjantsev},
  {Kim}, {Kim}, {Kim}, {Kim}, {King}, {Kishimoto}, {Kobayashi}, {Koga},
  {Konaka}, {Kormos}, {Koshio}, {Korzenev}, {Kowalik}, {Kropp}, {Kudenko},
  {Kurjata}, {Kutter}, {Kuze}, {Labarga}, {Lagoda}, {Lasorak}, {Laveder},
  {Lawe}, {Learned}, {Lim}, {Lindner}, {Litchfield}, {Longhin}, {Loverre},
  {Lou}, {Ludovici}, {Ma}, {Magaletti}, {Mahn}, {Malek}, {Maret}, {Mariani},
  {Martens}, {Marti}, {Martin}, {Marzec}, {Matsuno}, {Mazzucato}, {McCarthy},
  {McCauley}, {McFarland}, {McGrew}, {Mefodiev}, {Mermod}, {Metelko},
  {Mezzetto}, {Migenda}, {Mijakowski}, {Minakata}, {Minamino}, {Mine},
  {Mineev}, {Mitra}, {Miura}, {Mochizuki}, {Monroe}, {Moon}, {Moriyama},
  {Mueller}, {Muheim}, {Murase}, {Muto}, {Nakahata}, {Nakajima}, {Nakamura},
  {Nakaya}, {Nakayama}, {Nantais}, {Needham}, {Nicholls}, {Nishimura}, {Noah},
  {Nova}, {Nowak}, {Nunokawa}, {Obayashi}, {O'Keeffe}, {Okajima}, {Okumura},
  {Onishchuk}, {O'Sullivan}, and {O'Sullivan}]{hyperK2018}
{Hyper-Kamiokande Proto-Collaboration}; {Abe}, K.; {Abe}, K.; {Aihara},
  H.; {Aimi}, A.; {Akutsu}, R.; {Andreopoulos}, C.; {Anghel}, I.; {Anthony},
  L.H.V.;  Aushev, V.; et~al.
\newblock {Hyper-Kamiokande Design Report}.
\newblock {\em arXiv } {\bf 2018}, arXiv:1805.04163.
\newblock {\url{https://doi.org/10.48550/arXiv.1805.04163}}.

\bibitem[{The KM3NeT Collaboration} et~al.(2024){The KM3NeT Collaboration},
  {Aiello}, {Albert}, {Alhebsi}, {Alshamsi}, {Alves Garre}, {Ambrosone},
  {Ameli}, {Andre}, {Aphecetche}, {Ardid}, {Ardid}, {Atmani}, {Aublin},
  {Badaracco}, {Bailly-Salins}, {Barda{\v{c}}ov{\'a}}, {Baret},
  {Bariego-Quintana}, {Becherini}, {Bendahman}, {Benfenati}, {Benhassi},
  {Bennani}, {Benoit}, {Berbee}, {Bertin}, {Biagi}, {Boettcher}, {Bonanno},
  {Bouasla}, {Boumaaza}, {Bouta}, {Bouwhuis}, {Bozza}, {Bozza},
  {Br{\^a}nza{\c{s}}}, {Bretaudeau}, {Breuhaus}, {Bruijn}, {Brunner}, {Bruno},
  {Buis}, {Buompane}, {Busto}, {Caiffi}, {Calvo}, {Capone}, {Carenini},
  {Carretero}, {Cartraud}, {Castaldi}, {Cecchini}, {Celli}, {Cerisy}, {Chabab},
  {Chen}, {Cherubini}, {Chiarusi}, {Circella}, {Cocimano}, {Coelho}, {Coleiro},
  {Condorelli}, {Coniglione}, {Coyle}, {Creusot}, {Cuttone}, {Dallier}, {De
  Benedittis}, {De Martino}, {De Wasseige}, {Decoene}, {Del Rosso}, {Di Mauro},
  {Di Palma}, {D{\'\i}az}, {Diego-Tortosa}, {Distefano}, {Domi}, {Donzaud},
  {Dornic}, {Drakopoulou}, {Drouhin}, {Ducoin}, {Dvornick{\'y}}, {Eberl},
  {Eckerov{\'a}}, {Eddymaoui}, {van Eeden}, {Eff}, {van Eijk}, {El Bojaddaini},
  {El Hedri}, {Ellajosyula}, {Enzenh{\"o}fer}, {Ferrara}, {Filipovi{\'c}},
  {Filippini}, {Franciotti}, {Fusco}, {Gagliardini}, {Gal}, {Garc{\'\i}a
  M{\'e}ndez}, {Garcia Soto}, {Gatius Oliver}, {Gei{\ss}elbrecht}, {Genton},
  {Ghaddari}, {Gialanella}, {Gibson}, {Giorgio}, {Goos}, {Goswami}, {Gozzini},
  {Gracia}, {Guidi}, {Guillon}, {Guti{\'e}rrez}, {Haack}, {van Haren},
  {Heijboer}, {Hennig}, {Hern{\'a}ndez-Rey}, {Idrissi Ibnsalih}, {Illuminati},
  {Joly}, {de Jong}, {de Jong}, {Jung}, {Kistauri}, {Kopper}, {Kouchner},
  {Kovalev}, {Kueviakoe}, {Kulikovskiy}, {Kvatadze}, {Labalme}, {Lahmann},
  {Lamoureux}, {Larosa}, {Lastoria}, {Lazo}, {Le Stum}, {Lehaut},
  {Lema{\^\i}tre}, {Leonora}, {Lessing}, {Levi}, {Lindsey Clark}, {Longhitano},
  {Magnani}, {Majumdar}, {Malerba}, {Mamedov}, {Ma{\'n}czak}, {Manfreda},
  {Marconi}, {Margiotta}, {Marinelli}, {Markou}, {Martin}, {Mastrodicasa},
  {Mastroianni}, {Mauro}, {Miele}, {Migliozzi}, {Migneco}, {Mitsou}, {Mollo},
  {Morales-Gallegos}, {Moussa}, {Mozun Mateo}, {Muller}, {Musone}, {Musumeci},
  {Navas}, {Nayerhoda}, {Nicolau}, {Nkosi}, {Fearraigh}, {Oliviero}, {Orlando},
  {Oukacha}, {Paesani}, {Palacios Gonz{\'a}lez}, {Papalashvili}, {Parisi},
  {Pastor Gomez}, {P{\u{a}}un}, {P{\u{a}}v{\u{a}}la{\c{s}}}, {Pe{\~n}a
  Mart{\'\i}nez}, {Perrin-Terrin}, {Pestel}, {Pestes}, {Piattelli}, {Plavin},
  {Poir{\`e}}, {Popa}, and {Pradier}]{ORCA2024}
{The KM3NeT Collaboration}; {Aiello}, S.; {Albert}, A.; {Alhebsi}, A.R.;
  {Alshamsi}, M.; {Alves Garre}, S.; {Ambrosone}, A.; {Ameli}, F.; {Andre}, M.;
  {Aphecetche}, L.;  et~al.
\newblock {Measurement of neutrino oscillation parameters with the first six
  detection units of KM3NeT/ORCA}.
\newblock {\em J. High Energy Phys.} {\bf 2024}, {\em 2024},~206.
\newblock {\url{https://doi.org/10.1007/JHEP10(2024)206}}.

\bibitem[{Sanders}(2013)]{TMT2013}
{Sanders}, G.H.
\newblock {The Thirty Meter Telescope (TMT): An International Observatory}.
\newblock {\em J. Astrophys. Astron.} {\bf 2013}, {\em
  34},~81--86.
\newblock {\url{https://doi.org/10.1007/s12036-013-9169-5}}.

\bibitem[{Neichel} et~al.(2018){Neichel}, {Mouillet}, {Gendron}, {Correia},
  {Sauvage}, and {Fusco}]{EELT2018}
{Neichel}, B.; {Mouillet}, D.; {Gendron}, E.; {Correia}, C.; {Sauvage}, J.F.;
  {Fusco}, T.
\newblock {Overview of the European Extremely Large Telescope and its
  instrument suite}.
\newblock In Proceedings of the SF2A-2018: Proceedings of the Annual Meeting of
  the French Society of Astronomy and Astrophysics, {Bordeaux, France, 3--6 July 2018;} 
\newblock {\url{https://doi.org/10.48550/arXiv.1812.06639}}.

\bibitem[{Padovani} and {Cirasuolo}(2023)]{ELT2023}
{Padovani}, P.; {Cirasuolo}, M.
\newblock {The Extremely Large Telescope}.
\newblock {\em Contemp. Phys.} {\bf 2023}, {\em 64},~47--64.
\newblock {\url{https://doi.org/10.1080/00107514.2023.2266921}}.

\bibitem[{Barcons} et~al.(2012){Barcons}, {Barret}, {Decourchelle}, {den
  Herder}, {Dotani}, {Fabian}, {Fraga-Encinas}, {Kunieda}, {Lumb}, {Matt},
  {Nandra}, {Piro}, {Rando}, {Sciortino}, {Smith}, {Str{\"u}der}, {Watson},
  {White}, and {Willingale}]{Athena2012}
{Barcons}, X.; {Barret}, D.; {Decourchelle}, A.; {den Herder}, J.W.; {Dotani},
  T.; {Fabian}, A.C.; {Fraga-Encinas}, R.; {Kunieda}, H.; {Lumb}, D.; {Matt},
  G.;  et~al.
\newblock {Athena (Advanced Telescope for High ENergy Astrophysics) Assessment
  Study Report for ESA Cosmic Vision 2015--2025}.
\newblock {\em arXiv } {\bf 2012},  arXiv:1207.2745.
\newblock {\url{https://doi.org/10.48550/arXiv.1207.2745}}.

\bibitem[{Barret} et~al.(2020){Barret}, {Decourchelle}, {Fabian}, {Guainazzi},
  {Nandra}, {Smith}, and {den Herder}]{Athena2020}
{Barret}, D.; {Decourchelle}, A.; {Fabian}, A.; {Guainazzi}, M.; {Nandra}, K.;
  {Smith}, R.; {den Herder}, J.W.
\newblock {The \textit{Athena} space X‑ray observatory and the astrophysics of hot
  plasma{\textdagger}}.
\newblock {\em Astron.  Nachrichten} {\bf 2020}, {\em 341},~224--235.
\newblock {\url{https://doi.org/10.1002/asna.202023782}}.

\bibitem[{XRISM Science Team}(2020)]{xrism2020}
{XRISM Science Team}.
\newblock {Science with the X-ray Imaging and Spectroscopy Mission (XRISM)}.
\newblock {\em arXiv} {\bf 2020}, arXiv:2003.04962.
\newblock {\url{https://doi.org/10.48550/arXiv.2003.04962}}.

\bibitem[{Ajello} et~al.(2022){Ajello}, {Baldini}, {Ballet}, {Bastieri},
  {Becerra Gonzalez}, {Bellazzini}, {Berretta}, {Bissaldi}, {Bonino}, {Brill},
  {Bruel}, {Buson}, {Caputo}, {Caraveo}, {Cheung}, {Chiaro}, {Cibrario},
  {Ciprini}, {Crnogorcevic}, {Cutini}, {D'Ammando}, {De Gaetano}, {Di Lalla},
  {Di Venere}, {Dom{\'\i}nguez}, {Ramazani}, {Ferrara}, {Fiori}, {Fukazawa},
  {Funk}, {Fusco}, {Gammaldi}, {Gargano}, {Garrappa}, {Gasparrini},
  {Giglietto}, {Giordano}, {Giroletti}, {Green}, {Grenier}, {Guiriec}, {Horan},
  {Hou}, {Kayanoki}, {Kuss}, {Larsson}, {Latronico}, {Lewis}, {Li}, {Liodakis},
  {Longo}, {Loparco}, {Lott}, {Lovellette}, {Lubrano}, {Madejski}, {Maldera},
  {Manfreda}, {Mart{\'\i}-Devesa}, {Mazziotta}, {Mereu}, {Michelson},
  {Mirabal}, {Mitthumsiri}, {Mizuno}, {Monzani}, {Morselli}, {Moskalenko},
  {Negro}, {Ojha}, {Orienti}, {Orlando}, {Ormes}, {Pei}, {Pe{\~n}a-Herazo},
  {Persic}, {Pesce-Rollins}, {Petrosian}, {Pillera}, {Poon}, {Porter},
  {Principe}, {Rain{\`o}}, {Rando}, {Rani}, {Razzano}, {Razzaque}, {Reimer},
  {Reimer}, {Scotton}, {Serini}, {Sgr{\`o}}, {Siskind}, {Spandre}, {Spinelli},
  {Suson}, {Tajima}, {Torres}, {Valverde}, {Yassin}, and
  {Zaharijas}]{Fermi2022}
{Ajello}, M.; {Baldini}, L.; {Ballet}, J.; {Bastieri}, D.; {Becerra Gonzalez},
  J.; {Bellazzini}, R.; {Berretta}, A.; {Bissaldi}, E.; {Bonino}, R.; {Brill},
  A.;  et~al.
\newblock {The Fourth Catalog of Active Galactic Nuclei Detected by the Fermi
  Large Area Telescope: Data Release 3}.
\newblock {\em Astrophys. J. Suppl. Ser. } {\bf 2022}, {\em 263},~24.
\newblock {\url{https://doi.org/10.3847/1538-4365/ac9523}}.

\end{thebibliography}
\end{document}